\documentclass[numeric,12p,3p]{elsarticle}

 \usepackage{graphics}
 \usepackage{graphicx}
\usepackage{caption}
\usepackage{subcaption}
\usepackage{amssymb}
\usepackage{amsmath}
\usepackage{color}

\newtheorem{definition}{Definition}

\newtheorem{lemma}{Lemma}

\usepackage{hyperref}
\hypersetup{
colorlinks=true
}
\usepackage{epstopdf}
\usepackage{float} 

\biboptions{authoryear}

\journal{Journal of Time Series Analysis}

\begin{document}

\begin{frontmatter}



\title{
Dependence structure for the product of bi-dimensional finite-variance VAR(1) model components. An application to the cost of electricity load prediction errors.
}


\author[label2]{Joanna Janczura}
\author[label2]{Andrzej Pu\'c}
 \author[label1]{{\L}ukasz Bielak}
\author[label2]{Agnieszka Wy\l oma\'nska}
\address[label2]{Faculty of Pure and Applied Mathematics, Hugo Steinhaus Center, Wroclaw University of Science and Technology, Wyspia\'nskiego 27, 50-370 Wroc\l aw, Poland}
\address[label1]{KGHM, M. Sk\l odowskiej-Curie 48, 59-301 Lubin, Poland}

\begin{abstract}
In this paper we analyze the product of bi-dimensional VAR(1) model components. For the introduced time series we derive general formulas for the autocovariance function and study its properties for different cases of cross-dependence between the VAR(1) model components. The theoretical results are then illustrated in the simulation study for two types of bivariate distributions of the residual series, namely the Gaussian and Student's t. We also show a possible practical application of the obtained results based on the data from the electricity market.    
\end{abstract}


\end{frontmatter}
\section{Introduction}
One of important branches of probability theory is the analysis of a product of two (or even more) random variables. This issue is also relevant for statistics and applied mathematics. When one considers the product of random variables, the main attention is paid to its distribution and the analysis on how the probabilistic properties of the individual random variables influence the characteristics of their product. In the literature one can find research devoted to the analysis when the marginal random variables come from the same class of distributions with the special attention on the Gaussian and Student's t cases \citep[e.g. ][]{Ahsanullah_2014,student_t,gauss2,gauss1}. See also \cite{logistic,pearson,eliptically,trapezoidal,exponential,dirichlet,beta,beta2,pareto2}.
However,  one can also find the analysis related to the product of random variables coming from different classes of distributions, see e.g.  Gaussian and Laplace distributions \citep{normal_laplace,gauss_laplace2}, Gamma and Weibull distributions \citep{gama_weibul}, Gamma and Beta distributions \citep{gama_beta} or Pareto and Gamma distributions \citep{pareto_gama}.  For other references we refer the readers to \cite{maxwel,pearson_2,pareto4}. The theoretical results related to the product random variables were used in various applications, including finance, risk management, economy, but also  physical sciences, reliability theory, hydrology, and many others, see e.g.  \cite{appl1,appl2,appl3,appl4,appl5,appl6,appl7,appl8,appl10}. 

In this paper we extend the methodology related to the product random variables and analyze the product of two time series being the components of bi-dimensional discrete-time model, namely the vector autoregressive time series of order $1$,  called VAR(1) \citep[see e.g. ][]{brockwell2016introduction}. VAR is one of the classical multi-dimensional models used in various applications, especially in economy and finance \citep{econ1,econ2,fin1}. There are also many interesting research studies devoted to the theoretical analysis of VAR time series, see e.g. \cite{lut1,lut2}.

To our best knowledge, the problem related to time series (or general stochastic process) being a product of other time series is rarely discussed in the literature, but from the theoretical as well as practical point of view it seems to be very interesting. We refer the readers to the paper of \cite{wecker}, where the basic statistics are discussed for time series being a product of two stationary models. In the mentioned bibliography position, the author highlighted that the product time series is crucial in nonlinear time series analysis or in the theory of time series with random parameters. An analysis related to the product of stochastic (time dependent) components is presented by \cite{tella}, where the authors discuss the product of stochastic iterated integrals associated with general L\'evy processes. See also \cite{lee2004product,russo1998product}.

In the theory of finite-variance  time series, the most important characteristics describing the model is the autocovariance function (ACVF) in contrast to the theory of random variables, where the crucial point is the probabilistic distribution. Thus, in this paper, the main attention will be devoted to the analysis of the mean function and the ACVF of the time series being a product of two components of the finite-variance VAR(1) model. We check how the parameters of the bi-dimensional model influence the product's characteristics. The special attention is paid to the correlation coefficients of the residual series. We will show that the dependence between the residual series components of the VAR(1) model have a strong impact on the ACVF of the analysed time series independently of the residuals distribution. The distribution of the product time series is not deeply analyzed in this paper, however it is worth mentioning, that it can be obtained by using the results for the product random variables, see e.g. \cite{nasza_produkt}.   

In this paper, we also analyze special cases related to the dependency between components of the bi-dimensional model and discuss how their cross-dependence structure influences the ACVF of the product time series. Two cases of the residuals' distribution are examined in the simulation study, namely the bivariate Gaussian and bivariate Student's t and their impact on the analyzed model is demonstrated. The presented simulation studies indicate the important differences between the cases related to the dependence of the VAR(1) model's components and their distribution. 

Finally, the theoretical results are applied to the Danish electricity market case study. We show that the total cost of load prediction errors, which is the product of these errors and electricity prices, can be well described by time series being the product of the VAR(1) model components with the Student's t distribution. The presented approach yields a model that is consistent for both variables as well as their product.  

The rest of the paper is organized as follows. In Section \ref{sec:model} we recall the definition and main properties of the bi-dimensional VAR(1) model. Next, in Section \ref{sec:theory} we introduce the product time series and derive general formulas for its mean and autocovariance function. Special cases are then analyzed theoretically in Section \ref{special_cases} and illustrated using simulated data in Section \ref{sec:simulations}. Finally, Section \ref{sec:case_study} shows a possible practical application of the obtained results and Section \ref{sec:conclusions} concludes the paper.

\section{The bi-dimensional finite-variance VAR(1) model} \label{sec:model}
Let us first remind the definition and main properties of the bi-dimensional vector autoregressive time series of order 1, called VAR(1) model. 
\begin{definition}\label{def1}\citep{brockwell2016introduction} The bi-dimensional finite-variance VAR(1) time series  $\{\textbf{X}(t),~t\in \mathbb{Z}\}$ satisfies the following equation
\begin{eqnarray}\label{for1}
\textbf{X}(t)-\Phi\textbf{X}(t-1)=\textbf{Z}(t),
\end{eqnarray}
where $\textbf{X}(t)=(X_1(t),X_2(t))$, ${\Phi}$ is $2\times 2$ matrix
\begin{equation}\label{for2}
{\Phi}=\left[
\begin{array}{cccc}
\phi_{11} & \phi_{12}\\
\phi_{21}&\phi_{22}\\
\end{array}
\right],
\end{equation}
and $\{\textbf{Z}(t),~t\in \mathbb{Z}\}$ is the zero-mean bi-dimensional residual series, i.e. for each $t\in \mathbb{Z}$,  $\mathbf{Z}(t)=(Z_1(t),Z_2(t))$.
\end{definition}
In this paper we assume that $\{\textbf{Z}(t)\}$ is a series of independent bi-dimensional random variables having the same distribution, i.e. for each  $t\in \mathbb{Z}$, $(Z_1(t),Z_2(t))\sim (Z_1,Z_2)$. Moreover, we consider only the finite-variance  case, i.e. the covariance matrix of  $(Z_1,Z_2)$ (denoted further as $\Gamma_Z$) is properly defined. In the further analysis the variances of random variables $Z_1$ and  $Z_2$ for any $t$ are denoted as $\sigma_{Z,1}^2$ and $\sigma_{Z,2}^2$, respectively, while the correlation coefficient between them as $\rho_Z$.  

Let us assume that the following condition for the model coefficients is satisfied
{\begin{equation}
\text{det}(I-z\Phi)\neq0 \quad \text{for all } z \in \mathbb{Z} \text{ such that}\ |z| \leq 1,
\label{condition}
\end{equation}}
i.e. the eigenvalues of the matrix $\Phi$ (denoted further as $\nu_1$ and $\nu_2$) are less than $1$ in the absolute value. Under this assumption, for each $t\in \mathbb{Z}$ one can express $\mathbf{X}(t)$ in the causal representation 
{\begin{align}
\mathbf{X}(t)=\sum_{j=0}^{+\infty}\Phi^j\mathbf{Z}\left(t-j\right).
\label{eq2}
\end{align}}
Let us note that, when condition (\ref{condition}) is satisfied, then the coefficients $\Phi^j$ are absolutely summable. In this case, the time series given in Eq. (\ref{eq2}) is the unique bounded stationary solution of Eq.  (\ref{for1}) and it converges \citep{brockwell2016introduction}. In this paper, we consider only the case when the eigenvalues of the matrix $\Phi$ are the real numbers. 

We take the following notation
\begin{equation}\label{for22}
{\Phi^j}=\left[
\begin{array}{cccc}
\phi_{11}^{(j)} & \phi_{12}^{(j)}\\
\phi_{21}^{(j)}&\phi_{22}^{(j)}\\
\end{array}
\right],~~j=0,2\ldots.
\end{equation}
Obviously, for $j=0$, $\phi_{11}^{(j)}=\phi_{22}^{(j)}=1$ and $\phi_{12}^{(j)}=\phi_{21}^{(j)}=0$. \cite{matrix} shown that for a $2 \times 2$ matrix, the coefficients of $\Phi^j$ can be expressed in the following form depending on the eigenvalues of the matrix $\Phi$:
\begin{itemize}
    \item if $\nu_1$, $\nu_2$ are different eigenvalues of the matrix $\Phi$, i.e. $(\phi_{11}-\phi_{22})^2 \neq -4\phi_{21}\phi_{12}$ (and $|\nu_1|<1$, $|\nu_2|<1$), then we have
\begin{equation}
	\Phi^j= \left[
	\begin{array}{cc}
	\frac{\nu_2 \nu_1^j-\nu_1 \nu_2^j}{\nu_2-\nu_1}+\frac{\nu_2^j-\nu_1^j}{\nu_2-\nu_1}\phi_{11} & \frac{\nu_2^j-\nu_1^j}{\nu_2-\nu_1}\phi_{12}\\
	\frac{\nu_2^j-\nu_1^j}{\nu_2-\nu_1}\phi_{21} & \frac{\nu_2 \nu_1^j-\nu_1 \nu_2^j}{\nu_2-\nu_1}+\frac{\nu_2^j-\nu_1^j}{\nu_2-\nu_1}\phi_{22}
	\end{array}
	\right],~~j=1,2,\ldots,
	\label{phi_j1}
\end{equation}
\item if the eigenvalues of the matrix $\Phi$ are equal $\nu_1=\nu_2=\nu$, i.e. $(\phi_{11}-\phi_{22})^2 = -4\phi_{21}\phi_{12}$ (and $|\nu|<1$), then we  have
\begin{equation}
	\Phi^j= \left[
	\begin{array}{cc}
	j\nu^{j-1}\phi_{11}-(j-1)\nu^j & j\nu^{j-1}\phi_{12}\\
	j\nu^{j-1}\phi_{21} & j\nu^{j-1}\phi_{22}-(j-1)\nu^j
	\end{array}
	\right],~j=1,2\ldots.
\label{phi_j2}
\end{equation}
\end{itemize}
Using Eq. (\ref{eq2}) one can show that the components of the VAR(1) model can be expressed in the following form
\begin{eqnarray}\label{rep1}
X_i(t)=\sum_{j=0}^{\infty}\sum_{k=1}^2\phi^{(j)}_{ik}Z_{k}(t-j),~~i=1,2
\end{eqnarray}
and their distributions do not depend on $t$. Thus, from Eq. (\ref{rep1}) one can obtain the  formulas for variances $\sigma_{X,1}^2$, $\sigma_{X,2}^2$ of $X_1(t)$ and $X_2(t)$, respectively 
\begin{eqnarray}\label{for11_new}
\sigma_{X,i}^2=\mathbb{V}\text{ar}(X_i(t))=\sum_{j=0}^{\infty}\sum_{k,l=1}^2\phi_{ik}^{(j)}\phi_{il}^{(j)}\gamma_{Z,k,l},
\end{eqnarray}
where $\gamma_{Z,k,l}$ is the $(k,l)$ component of the covariance matrix $\Gamma_Z$. Recall that $\Gamma_Z$ is is given by
\begin{eqnarray}
\Gamma_Z=\Gamma_{Z}(t)=\left[\gamma_{Z,i,j}(t)\right]_{i,j=1}^2=\left[\mathbb{E}[Z_{i}(t)Z_j(t)]\right]_{i,j=1}^2,
\end{eqnarray} 
where $\gamma_{Z,i,i}=\sigma_{Z,i}^2$ and $\gamma_{Z,1,2}=\gamma_{Z,2,1}=\rho_Z\sigma_{Z,1}\sigma_{Z,2}$.

Let us note that the covariance between  $X_1(t)$ and $X_2(t)$ is also independent on $t$ and it is given by
\begin{eqnarray}\label{cov_XX}
\gamma_{X,1,2}&=&\mathbb{E}[X_1(t)X_2(t)]=\mathbb{E}\Bigg[\sum_{j=0}^{\infty}\sum_{k=1}^2\phi^{(j)}_{1k}Z_{k}(t-j)\sum_{i=0}^{\infty}\sum_{l=1}^2\phi^{(i)}_{2l}Z_{l}(t-i)\Bigg]\nonumber\\
&=&\sum_{j=0}^{\infty}\sum_{k,l=1}^2\phi^{(j)}_{1k}\phi^{(j)}_{2l}\mathbb{E}[Z_{k}(t-j)Z_l(t-j)] =\sum_{j=0}^{\infty}\sum_{k,l=1}^2\phi^{(j)}_{1k}\phi^{(j)}_{2l}\gamma_{Z,k,l}.
\end{eqnarray}
Thus, the correlation coefficient between $X_1(t)$ and $X_2(t)$ for each $t \in \mathbb{Z}$ is given by
\begin{eqnarray}\label{cor_X}
\rho_{X}=\frac{\gamma_{X,1,2}}{\sigma_{X,1}\sigma_{X,2}}=\frac{\sum_{j=0}^{\infty}\sum_{k,l=1}^2\phi^{(j)}_{1k}\phi^{(j)}_{2l}\gamma_{Z,k,l}}{\sqrt{\sum_{j=0}^{\infty}\sum_{k,l=1}^2\phi_{1k}^{(j)}\phi_{1l}^{(j)}\gamma_{Z,k,l}\sum_{j=0}^{\infty}\sum_{k,l=1}^2\phi_{2k}^{(j)}\phi_{2l}^{(j)}\gamma_{Z,k,l}}}.
\end{eqnarray}
The autocovariance function of $\{X_i(t)\}$ for $i=1,2$ is independent on $t$ and takes the form
\begin{eqnarray}
 ACVF_{X_i}(h)&=&\mathbb{E}[X_i(t)X_i(t+h)]=\sum_{j=0}^{\infty}\sum_{k,l=1}^2\phi^{(j)}_{ik}\phi^{(h+j)}_{il}\gamma_{Z,k,l}.
\end{eqnarray}
Using the same reasoning as in the above calculations, one can show that the cross-covariance between $\{X_1(t)\}$ and $\{X_2(t)\}$ is also independent on $t$ and it is given by
\begin{eqnarray}\label{ccov_X}
CCVF_{X_1,X_2}(h)&=&\mathbb{E}[X_1(t)X_2(t+h)]=\sum_{j=0}^{\infty}\sum_{k,l=1}^2\phi^{(j)}_{1k}\phi^{(h+j)}_{2l}\gamma_{Z,k,l}.
\end{eqnarray}
\section{Product of the components of bi-dimensional finite-variance VAR(1) model} \label{sec:theory}

Here, we introduce the time series $\{Y(t),~t\in \mathbb{Z}\}$ that is a product of two components of the bi-dimensional VAR(1) model discussed in the previous section. Precisely, for each $t\in \mathbb{Z}$ we have
\begin{eqnarray}\label{for3}
Y(t)=X_1(t)X_2(t),
\end{eqnarray}
where the bi-dimensional time series $\{\mathbf{X}(t)\}$ satisfies  Eq. (\ref{for1}). Assuming that condition (\ref{condition}) is fulfilled and applying Eq. (\ref{rep1}) one can show that for each $t\in \mathbb{Z}$, $Y(t)$  can be represented as
\begin{eqnarray}\label{rep2}
Y(t)=\sum_{j,i=0}^{\infty}\sum_{k,l=1}^2\phi^{(j)}_{1k}\phi^{(i)}_{2l}Z_{k}(t-j)Z_{l}(t-i).
\end{eqnarray}
Using the above representation, one can calculate the main characteristics of the $\{Y(t)\}$ time series. In the lemmas presented below, we assume that in general, $\mathbb{E}[Z_kZ_lZ_nZ_r]<\infty$ for $k,l,n,r=1,2$. However, for some special cases (see Section  \ref{special_cases}), this assumption may be less restrictive depending on the relationship between components of the considered VAR(1) model.

\begin{lemma}\label{lem1}
If  $\{Y(t)\}$ is the product time series defined in Eq. (\ref{for3}), where $\{X_1(t)\}$ and $\{X_2(t)\}$ are the two components of the finite-variance VAR(1) model given in Definition  \ref{def1} satisfying the condition (\ref{condition}), then the expected value and variance of $\{Y(t)\}$ exist and are given by
\begin{eqnarray}\label{ex_varr}
\mathbb{E}(Y(t))&=&\gamma_{X,1,2}=\rho_X\sigma_{X,1}\sigma_{X,2},\\
\mathbb{V}\text{ar}(Y(t))&=&\sigma^2_Y=\mathbb{E}\left[\left(\sum_{j,i=0}^{\infty}\sum_{k,l=1}^2\phi^{(j)}_{1k}\phi^{(i)}_{2l}Z_{k}(t-j)Z_{l}(t-i)\right)^2\right]-\gamma_{X,1,2}^2,
\end{eqnarray}
where $\sigma_{X,i}$, $\gamma_{X,1,2}$, $\rho_X$ are given in Eqs. (\ref{for11_new}), (\ref{cov_XX}) and (\ref{cor_X}), respectively.
\end{lemma}
The proof of Lemma \ref{lem1} follows directly from Eqs. (\ref{for3}) and (\ref{rep2}).

In the following lemma, we present the formula for the  ACVF of the time series $\{Y(t)\}$,  ACVF$_Y(t,t+h)=\text{Cov}(Y(t),Y(t+h))$ for $t,h\in \mathbb{Z}$. 
\begin{lemma}\label{lem2}
If  $\{Y(t)\}$ is the product time series defined in Eq. (\ref{for3}), where $\{X_1(t)\}$ and $\{X_2(t)\}$ are the two components of the finite-variance VAR(1) model given in Definition  \ref{def1} satisfying the condition (\ref{condition}), then the autocovariance function of $\{Y(t)\}$ for $h=0,1,\ldots,$ exists  and it has the following form 
\begin{eqnarray}\label{acvfy}
\text{ACVF}_Y(t,t+h)&=&\sum_{j,i=0}^{\infty}\sum_{m,p=-h}^{\infty}\sum_{k,l,n,r=1}^2\phi^{(j)}_{1k}\phi^{(i)}_{2l}\phi^{(m+h)}_{1n}\phi^{(p+h)}_{2r}\mathbb{E}\left[Z_k(t-j)Z_l(t-i)Z_n(t-m)Z_r(t-p)\right]\nonumber \\
&&-\gamma_{X,1,2}^2,
\end{eqnarray}
where $\gamma_{X,1,2}$ is given in Eq. (\ref{cov_XX}).
\end{lemma}
\textit{Proof:} The expectations $\mathbb{E}(Y(t))$ and $\mathbb{E}(Y(t+h))$ are given in Eq. (\ref{ex_varr}). Thus, we need to calculate $\mathbb{E}(Y(t)Y(t+h))$ for any $h=0,1,\ldots$. Using Eq. (\ref{rep2}) one obtains
\begin{eqnarray*}
\mathbb{E}(Y(t)Y(t+h))
&=&\sum_{j,i,m,p=0}^{\infty}\sum_{k,l,n,r=1}^2\phi^{(j)}_{1k}\phi^{(i)}_{2l}\phi^{(m)}_{1n}\phi^{(p)}_{2r}\mathbb{E}\left[Z_k(t-j)Z_l(t-i)Z_n(t+h-m)Z_r(t+h-p)\right]\nonumber\\
&=&\sum_{j,i=0}^{\infty}\sum_{m,p=-h}^{\infty}\sum_{k,l,n,r=1}^2\phi^{(j)}_{1k}\phi^{(i)}_{2l}\phi^{(m+h)}_{1n}\phi^{(p+h)}_{2r}\mathbb{E}\left[Z_k(t-j)Z_l(t-i)Z_n(t-m)Z_r(t-p)\right].
\end{eqnarray*}
Applying the formula for the ACVF of $\{Y(t)\}$
\begin{eqnarray*}
\text{ACVF}_Y(t,t+h)=\mathbb{E}(Y(t)Y(t+h))-\mathbb{E}(Y(t))\mathbb{E}(Y(t+h))
\end{eqnarray*}
 we obtain the thesis.
\begin{flushright}
$\Box$
\end{flushright}

Let us emphasize that the ACVF given in Eq. (\ref{acvfy}) is independent on $t$. Moreover, $\{Y(t)\}$ has a constant mean function. Thus, it is stationary in the weak sense. Therefore, in the further analysis it will be denoted as ACVF$_Y(h)$.


\section{Special cases analysis}\label{special_cases}
In this section, we consider the following special cases related to the dependence of the components of the finite-variance VAR(1) model given in Definition \ref{def1}. Let us emphasise that in the considered cases we do not consider any specific distribution of the residual series. 
\begin{itemize}
    \item Case 1: the time series $\{X_1(t)\}$ and $\{X_2(t)\}$ are independent. This is the case, when for each $t\in \mathbb{Z}$ the random variables $Z_1$ and $Z_2$ are independent  and $\phi_{12}=\phi_{21}=0$, where $\phi_{ij}$ $i,j=1,2$ are the coefficients of the matrix $\Phi$ given in Eq. (\ref{for2}). In this case,  $\{X_1(t)\}$ and $\{X_2(t)\}$ are two independent autoregressive time series of order $1$ (called AR(1)) satisfying the following equations
    \begin{eqnarray}\label{ts_case1}
    X_1(t)-\phi_{11}X_1(t-1)=Z_1(t),~~X_2(t)-\phi_{22}X_2(t-1)=Z_2(t).
    \end{eqnarray}
    \item Case 2: the time series $\{X_1(t)\}$ and $\{X_2(t)\}$ are dependent only through the residual components. In this case, we assume that the random variables $Z_1$ and $Z_2$ are dependent (and we assume their correlation coefficient $\rho_{Z}\neq 0$), however $\phi_{12}=\phi_{21}=0$, where $\phi_{ij}$ $i,j=1,2$ are the coefficients of the matrix $\Phi$ given in Eq. (\ref{for2}). In this case,  $\{X_1(t)\}$ and $\{X_2(t)\}$ also satisfy Eq. (\ref{ts_case1}), but they are dependent. 
    \item Case 3: the time series $\{X_1(t)\}$ and $\{X_2(t)\}$ are dependent only through the coefficients of the VAR(1) model. This is the case, when the random variables $Z_1$ and $Z_2$ are independent, however $\phi_{12}\neq 0$ or/and  $\phi_{21}\neq 0$, where $\phi_{ij}$ $i,j=1,2$ are the coefficients of the matrix $\Phi$ given in Eq. (\ref{for2}).
    For simplicity we assume $\phi_{12}\neq 0$ and $\phi_{21}=0$. In this case,  $\{X_1(t)\}$ and $\{X_2(t)\}$ satisfy the following equations
    \begin{eqnarray}
    X_1(t)-\phi_{11}X_1(t-1)-\phi_{12}X_2(t-1)=Z_1(t),~~X_2(t)-\phi_{22}X_2(t-1)=Z_2(t),
    \end{eqnarray}
    thus, the time series $\{X_2(t)\}$ is the AR(1) model while $\{X_1(t)\}$ does not satisfy the AR(1) equation.
\end{itemize}
\subsection{Case 1}
In this case we assume $\sigma_{Z,i}^2<\infty$ for $i=1,2$. The coefficients of the matrix $\Phi$ given in Eq. (\ref{for2}) that lie outside the main diagonal are equal to zero, i.e., $\phi_{12}=\phi_{21}=0$. Thus, we have \begin{eqnarray}\label{case1}\phi_{11}^{(j)}=\phi_{11}^j, ~\phi_{22}^{(j)}=\phi_{22}^j,~ \phi_{12}^{(j)}=\phi_{21}^{(j)}=0,~j=0,1,\ldots .\end{eqnarray} 
Moreover, according to the condition given in Eq. (\ref{condition}), $|\phi_{11}|<1$ and $|\phi_{22}|<1$. Using Eq. (\ref{for11_new}) and (\ref{cor_X}) one can easily show that
\begin{eqnarray}\label{sigma_case2}
\sigma_{X,i}^2=\sum_{j=0}^{\infty}\phi_{ii}^{2j}\gamma_{Z,i,i}=\frac{\sigma_{Z,i}^2}{1-\phi_{ii}^2}.
\end{eqnarray}
In the considered case $\rho_X=0$.

Finally, using Eqs. (\ref{ex_varr}) and (\ref{acvfy}) one can  show that the following hold
\begin{equation}\label{ex_var_case2}
    \mathbb{E}(Y(t)) = 0,~~
    \mathbb{V}\text{ar}(Y(t)) = \frac{\sigma_{Z,1}^2\sigma_{Z,2}^2}{(1-\phi_{11}^2)(1-\phi_{22}^2)},~~
\text{ACVF}_Y(h)=\frac{\sigma_{Z,1}^2\sigma_{Z,2}^2(\phi_{11}\phi_{22})^h}{(1-\phi_{11}^2)(1-\phi_{22}^2)}.
\end{equation}

\subsection{Case 2}
In this case we assume $\mathbb{E}[Z_1^2Z_2^2]<\infty$. Similarly as previously, condition (\ref{case1}) is satisfied. However, now, we assume that the components of the residual series are dependent and the correlation coefficient $\rho_Z$  is non-zero.
One can show that $\sigma_{X,i}^2$ has the same form as in Case 1 for $i=1,2$, i.e., it is given by Eq. (\ref{sigma_case2}). However, using Eq. (\ref{cor_X}) we obtain that  the $\rho_X$ coefficient is given by
\begin{eqnarray}
\rho_{X}=\frac{\rho_Z\sqrt{(1-\phi_{11}^2)(1-\phi_{22}^2)}}{1-\phi_{11}\phi_{22}}.
\end{eqnarray}
Using Eq. (\ref{ex_varr}) one obtains
\begin{eqnarray}\label{ex_var_case33}
    \mathbb{E}(Y(t)) =\frac{\rho_Z\sigma_{Z,1}\sigma_{Z,2}}{(1-\phi_{11}\phi_{22})}.
    \end{eqnarray}
On the other hand, using Eq. (\ref{acvfy}) we can calculate the ACVF for $\{Y(t)\}$ for  $h=0,1,\dots$. Indeed, we have 
\begin{eqnarray*}
\text{ACVF}_Y(h)&=&\sum_{j,i=0}^{\infty}\sum_{m,p=-h}^{\infty}\phi^{j+m+h}_{11}\phi^{i+p+h}_{22}\mathbb{E}\left[Z_1(t-j)Z_2(t-i)Z_1(t-m)Z_2(t-p)\right]-\frac{\rho_Z^2\sigma_{Z,1}^2\sigma_{Z,2}^2}{(1-\phi_{11}\phi_{22})^2}.
\end{eqnarray*}
Now,  we can calculate the value $$r_{1,2}(t,j,m,i,p)=\mathbb{E}\left[Z_1(t-j)Z_1(t-m)Z_2(t-i)Z_2(t-p)\right]$$ for all $t\in \mathbb{Z}$ and $i,j=0,1,\dots$, $m,p=-h,-h+1,\dots$. Using the fact that for each $t\in \mathbb{Z}$ the bi-dimensional residual series  $\textbf{Z}(t)$ is a zero-mean vector and  for $t\neq s$, $\textbf{Z}(t)$ is independent on $\textbf{Z}(s)$, one obtains the following
\begin{equation}\label{EZ4}
r_{1,2}(t,j,m,i,p)= 
\begin{cases}
\mathbb{E}\left[Z_1^2(t-j)Z_2^2(t-j)\right], ~ \mbox{if} ~ i=j=p=m;\\\\
\mathbb{E}\left[Z_1^2(t-j)\right]\mathbb{E}\left[Z_2^2(t-i)\right], ~ \mbox{if} ~ j=m,i=p,j\neq i;\\\\
\mathbb{E}\left[Z_1(t-j)Z_2(t-j)\right]\mathbb{E}\left[Z_1(t-m)Z_2(t-m)\right], ~ \mbox{if} ~ j=i,m=p,j\neq m; \\\\
\mathbb{E}\left[Z_1(t-j)Z_2(t-j)\right]\mathbb{E}\left[Z_1(t-m)Z_2(t-m)\right],~ \mbox{if} ~ j=p,i=m,j\neq i.
\end{cases}
\end{equation}
Thus, we have
\begin{equation}\label{EZ44}
r_{1,2}(t,j,m,i,p)=
\begin{cases}
m_Z, ~ \mbox{if} ~ i=j=p=m;~i,j,m,p=0,1,2\ldots ;\\\\
\sigma_{Z,1}^2\sigma_{Z,2}^2, ~ \mbox{if} ~ j=m,i=p,j\neq i;~~i,j,m,p=0,1,2\ldots ;\\\\
\rho_{Z}^2\sigma_{Z,1}^2\sigma_{Z,2}^2, ~ \mbox{if} ~ j=i,m=p,j\neq m; ~ i,j=0,1,2\ldots,~~m,p= -h,-h+1\ldots ;\\\\
\rho_{Z}^2\sigma_{Z,1}^2\sigma_{Z,2}^2, ~ \mbox{if} ~ j=p,i=m,j\neq i;~i,j,m,p=0,1,2\ldots,
\end{cases}
\end{equation}
where the value $m_Z=\mathbb{E}\left[Z_1^2(t)Z_2^2(t)\right]$ is independent on $t$. Therefore, we have
\begin{eqnarray}\label{acvfy_final_case3}
\text{ACVF}_Y(h)&=&m_Z\sum_{j=0}^{\infty}\left(\phi_{11}\phi_{22}\right)^{2j+h}+\sigma_{Z,1}^2\sigma_{Z,2}^2\sum_{j=0}^{\infty}\phi_{11}^{2j+h}\left[\sum_{i=0}^{\infty}\phi_{22}^{2i+h}-\phi_{22}^{2j+h}\right]\nonumber\\
&&+\rho_{Z}^2\sigma_{Z,1}^2\sigma_{Z,2}^2\sum_{j=0}^{\infty}\phi_{11}^{j+h}\left[\phi_{22}^{j+h}\sum_{m=0}^{\infty}\left(\phi_{11}\phi_{22}\right)^m-\phi_{22}^{j+h}\left(\phi_{11}\phi_{22}\right)^j\right]\nonumber\\
&&+\rho_{Z}^2\sigma_{Z,1}^2\sigma_{Z,2}^2\sum_{j=0}^{\infty}\phi_{11}^{j+h}\left[\phi_{22}^{j+h}\sum_{m=-h}^{\infty}\left(\phi_{11}\phi_{22}\right)^m-\phi_{22}^{j+h}\left(\phi_{11}\phi_{22}\right)^j\right]
-\frac{\rho_Z^2\sigma_{Z,1}^2\sigma_{Z,2}^2}{(1-\phi_{11}\phi_{22})^2}\nonumber\\
&=&\frac{m_Z\left(\phi_{11}\phi_{22}\right)^h}{1-\left(\phi_{11}\phi_{22}\right)^2}+\sigma_{Z,1}^2\sigma_{Z,2}^2\left(\phi_{11}\phi_{22}\right)^h\left(\frac{1}{(1-\phi_{11}^2)(1-\phi_{22}^2)}-\frac{1}{1-\left(\phi_{11}\phi_{22}\right)^{2}}\right)\nonumber\\
&&+\rho_{Z}^2\sigma_{Z,1}^2\sigma_{Z,2}^2\left(\phi_{11}\phi_{22}\right)^h\left(\frac{1}{\left(1-\phi_{11}\phi_{22}\right)^2}-\frac{1}{1-\left(\phi_{11}\phi_{22}\right)^2}\right)\nonumber\\
&&+\rho_{Z}^2\sigma_{Z,1}^2\sigma_{Z,2}^2\left(\phi_{11}\phi_{22}\right)^h\left(\frac{\left(\phi_{11}\phi_{22}\right)^{-h}}{\left(1-\phi_{11}\phi_{22}\right)^2}-\frac{1}{1-\left(\phi_{11}\phi_{22}\right)^2}\right)-\frac{\rho_Z^2\sigma_{Z,1}^2\sigma_{Z,2}^2}{(1-\phi_{11}\phi_{22})^2}\\
&=&\left(\phi_{11}\phi_{22}\right)^h\left[\frac{m_Z-\sigma_{Z,1}^2\sigma_{Z,2}^2-2\rho_Z^2\sigma_{Z,1}^2\sigma_{Z,2}^2}{1-\left(\phi_{11}\phi_{22}\right)^2}+\frac{\sigma_{Z,1}^2\sigma_{Z,2}^2}{(1-\phi_{11}^2)(1-\phi_{22}^2)}+\frac{\rho_Z^2\sigma_{Z,1}^2\sigma_{Z,2}^2}{\left(1-\phi_{11}\phi_{22}\right)^2}\right].\nonumber
\end{eqnarray}
Finally, taking $h=0$ one obtains that the variance of a random variable $Y(t)$ for each $t\in \mathbb{Z}$: 
\begin{equation*}
\mathbb{V}\text{ar}(Y(t))=\frac{m_Z-\sigma_{Z,1}^2\sigma_{Z,2}^2-2\rho_Z^2\sigma_{Z,1}^2\sigma_{Z,2}^2}{1-\left(\phi_{11}\phi_{22}\right)^2}+\frac{\sigma_{Z,1}^2\sigma_{Z,2}^2}{(1-\phi_{11}^2)(1-\phi_{22}^2)}+\frac{\rho_Z^2\sigma_{Z,1}^2\sigma_{Z,2}^2}{\left(1-\phi_{11}\phi_{22}\right)^2}.
\end{equation*}

\subsection{Case 3}
In this case we assume $\mathbb{E}(Z_1^2Z_2^2)<\infty$ and $\mathbb{E}(Z_2^4)<\infty$. One can show that we have \begin{eqnarray}\label{case3}\phi_{11}^{(j)}=\phi_{11}^j, ~\phi_{22}^{(j)}=\phi_{22}^j,~ \phi_{21}^{(j)}=0,~~~j=0,1,\ldots\end{eqnarray}
and the eigenvalues of the matrix $\Phi$  are equal $\nu_1=\phi_{11}$ and $\nu_2=\phi_{22}$. Thus, according to Eqs. (\ref{phi_j1}) and (\ref{phi_j2})  the following is fulfilled for $j=1,2,\ldots$ \begin{equation}\label{phi_12}
\phi_{12}^{(j)}= 
\begin{cases}
\frac{\phi_{22}^j-\phi_{11}^j}{\phi_{22}-\phi_{11}}\phi_{12}, ~ \mbox{if} ~ \phi_{11}\neq \phi_{22};\\\\
j\phi_{11}^{j-1}\phi_{12},~ \mbox{if} ~ \phi_{11}=\phi_{22},
\end{cases}
\end{equation}
while for $j=0$, $\phi_{12}^{(j)}=0$. In order to fulfill condition (\ref{condition}) we assume that  $|\phi_{11}|<1$ and $|\phi_{22}|<2$.

Using Eq. (\ref{for11_new}) one obtains
\begin{eqnarray}
\sigma_{X,1}^2=\frac{\sigma_{Z,1}^2}{1-\phi_{11}^2}+\sigma_{Z,2}^2\sum_{j=0}^{\infty}\left(\phi_{12}^{(j)}\right)^2,~~\sigma_{X,2}^2=\frac{\sigma_{Z,2}^2}{1-\phi_{22}^2}.
\end{eqnarray}
\noindent Thus, we have 
\begin{equation}
\sigma_{X,1}^2= 
\begin{cases}
\frac{\sigma_{Z,1}^2}{1-\phi_{11}^2}+\frac{\sigma_{Z,2}^2\phi_{12}^2}{(\phi_{22}-\phi_{11})^2}\left[\frac{\phi_{22}^2}{1-\phi_{22}^2}-\frac{2\phi_{11}\phi_{22}}{1-\phi_{11}\phi_{22}}+\frac{\phi_{11}^2}{1-\phi_{21}^2}\right], ~ \mbox{if} ~ \phi_{11}\neq \phi_{22};\\\\
\frac{\sigma_{Z,1}^2}{1-\phi_{11}^2}+\frac{\sigma_{Z,2}^2\phi_{12}^2(1+\phi_{11}^2)}{(1-\phi_{11}^2)^3},~ \mbox{if} ~ \phi_{11}=\phi_{22}.
\end{cases}
\end{equation}
Moreover from Eq. (\ref{cov_XX}) we have
\begin{eqnarray*}\label{cov_XX_new}
\gamma_{X,1,2}=\sum_{j=1}^{\infty}\sigma_{Z,2}^2\phi_{12}^{(j)}\phi_{22}^j.
\end{eqnarray*}
Thus, we obtain the following formula for the expected value of the random variable $Y(t)$ for each $t\in \mathbb{Z}$
\begin{equation}
\mathbb{E}(Y(t))=\gamma_{X,1,2}= 
\begin{cases}
\frac{\sigma_{Z,2}^2\phi_{12}}{\phi_{22}-\phi_{11}}\left[\frac{\phi_{22}^2}{1-\phi_{22}^2}-\frac{\phi_{11}\phi_{22}}{1-\phi_{22}\phi_{11}}\right], ~ \mbox{if} ~ \phi_{11}\neq \phi_{22};\\\\
\frac{\sigma_{Z,2}^2\phi_{22}\phi_{12}}{(1-\phi_{11}\phi_{22})^2},~ \mbox{if} ~ \phi_{11}=\phi_{22}.
\end{cases}
\end{equation}
To obtain the explicit formula for ACVF$_Y(h)$ we will use Eq. (\ref{acvfy}). For $h=0,1,2\ldots$ we have 
\begin{eqnarray*}
\text{ACVF}_Y(h)&=&\sum_{j,i=0}^{\infty}\sum_{m,p=-h}^{\infty}\sum_{k,l,n,r=1}^2\phi^{(j)}_{1k}\phi^{(i)}_{2l}\phi^{(m+h)}_{1n}\phi^{(p+h)}_{2r}\mathbb{E}\left[Z_k(t-j)Z_l(t-i)Z_n(t-m)Z_r(t-p)\right]-\gamma_{X,1,2}^2\\
&=&\sum_{j,i=0}^{\infty}\sum_{m,p=-h}^{\infty}\sum_{k,n=1}^2\phi^{(j)}_{1k}\phi^{(i)}_{22}\phi^{(m+h)}_{1n}\phi^{(p+h)}_{22}\mathbb{E}\left[Z_k(t-j)Z_2(t-i)Z_n(t-m)Z_2(t-p)\right]-\gamma_{X,1,2}^2\\
&=&\sum_{j,i=0}^{\infty}\sum_{m,p=-h}^{\infty}\phi^{i+p+h}_{22}\sum_{k,n=1}^2\phi^{(j)}_{1k}\phi^{(m+h)}_{1n}\mathbb{E}\left[Z_k(t-j)Z_2(t-i)Z_n(t-m)Z_2(t-p)\right]-\gamma_{X,1,2}^2\\
&=&\sum_{j,i=0}^{\infty}\sum_{m,p=-h}^{\infty}\phi^{i+p+h}_{22}\phi^{(j)}_{12}\phi^{(m+h)}_{12}\mathbb{E}\left[Z_2(t-j)Z_2(t-i)Z_2(t-m)Z_2(t-p)\right]\\
&+&\sum_{j,i=0}^{\infty}\sum_{m,p=-h}^{\infty}\phi^{i+p+h}_{22}\phi^{j+m+h}_{11}\mathbb{E}\left[Z_1(t-j)Z_2(t-i)Z_1(t-m)Z_2(t-p)\right]-\gamma_{X,1,2}^2.
\end{eqnarray*}
Moreover, the value
$$r_{2,2}(t,j,m,i,p)=\mathbb{E}\left[Z_2(t-j)Z_2(t-i)Z_2(t-m)Z_2(t-p)\right]$$
is given by
\begin{equation}
r_{2,2}(t,j,m,i,p)=
\begin{cases}
\kappa_Z, ~ \mbox{if} ~ i=j=m=p;~i,j,m,p=0,1,\ldots,\\\\
\sigma_{Z,2}^4,~ \mbox{if} ~ i=j,~~ m=p,~m\neq j;~i,j=0,1,\ldots,~m,p=-h,-h+1,\ldots,\\\\
\sigma_{Z,2}^4,~ \mbox{if} ~ j=m,~~ i=p,~m\neq i;~i,j,m,p=0,1,\ldots,\\\\
\sigma_{Z,2}^4,~ \mbox{if} ~ j=p,~~ i=m,~m\neq j;~i,j,m,p=0,1,\ldots,\\\\
\end{cases}
\end{equation}
where $\kappa_Z=\mathbb{E}\left[Z_2^4(t)\right]<\infty$ is independent of $t$. Additionally, one can show that
\begin{eqnarray*}
\mathbb{E}\left[Z_1(t-j)Z_2(t-i)Z_1(t-m)Z_2(t-p)\right]=\sigma_{Z,1}^2\sigma_{Z,2}^2,~\mbox{if}~j=m, i=p,~i,j,p,m = 0,1,2,\ldots.
\end{eqnarray*}
Thus, we have the following
\begin{eqnarray*}\label{acvfy11}
\text{ACVF}_Y(h)&=&\kappa_{Z}\sum_{j=0}^{\infty}\phi^{(j)}_{12}\phi^{2j+h}_{22}\phi^{(j+h)}_{12}+\sigma_{Z,2}^4\sum_{j=0}^{\infty}\phi_{12}^{(j)}\phi_{12}^{(j+h)}\left[\sum_{i=0}^{\infty}\phi_{22}^{2i+h}-\phi_{22}^{2j+h}\right]\\&&+\sigma_{Z,2}^4\sum_{j=0}^{\infty}\phi_{22}^{j+h}\phi_{12}^{(j)}\left[\sum_{i=0}^{\infty}\phi_{22}^i\phi_{12}^{(i+h)}-\phi_{22}^j\phi_{12}^{(j+h)}\right]\nonumber\\
&&+\sigma_{Z,2}^4\sum_{j=0}^{\infty}\phi_{22}^{j+h}\phi_{12}^{(j)}\left[\sum_{m=-h}^{\infty}\phi_{22}^m\phi_{12}^{(m+h)}-\phi_{22}^j\phi_{12}^{(j+h)}\right]\\
&&+\sigma_{Z,1}^2\sigma_{Z,2}^2\sum_{j=0}^{\infty}\sum_{i=0}^{\infty}\phi_{11}^{2j+h}\phi_{22}^{2i+h}-\gamma_{X,1,2}^2.
\end{eqnarray*}
Let us observe that for $h=0,1,2,\ldots$ the following holds
\begin{eqnarray}
\sum_{j=0}^{\infty}\sum_{i=0}^{\infty}\phi_{11}^{2j+h}\phi_{22}^{2i+h}=\frac{\left(\phi_{11}\phi_{22}\right)^h}{(1-\phi_{11}^2)(1-\phi_{22}^2)}.
\end{eqnarray}
Now, to make the calculations simpler, let us assume that $\phi_{11}=0$ and $\phi_{22}\neq 0$. In this case, the matrix $\Phi$ (see Eq. (\ref{for2})) has two different eigenvalues and $\phi_{12}^{(j)}=\phi_{12}\phi_{22}^{j-1}$, $j=1,2\ldots$. Clearly, $\phi_{11}^{(j)}=0$ for $j=1,2,\ldots$ and for $j=0$, $\phi_{11}^{j}=1$. We have the following
\begin{eqnarray*}
\sum_{j=0}^{\infty}\phi^{(j)}_{12}\phi^{2j+h}_{22}\phi^{(j+h)}_{12}&=&\phi_{12}^{2}\phi_{22}^{2h}\sum_{j=1}^{\infty}\phi_{22}^{4j-2}=\frac{\phi_{12}^{2}\phi_{22}^{2h}\phi_{22}^2}{1-\phi_{22}^4},~h=0,1,2,\ldots.
\end{eqnarray*}
Let us first consider the case $h=0$. We have
\begin{eqnarray*}
\sum_{j=0}^{\infty}\phi_{12}^{(j)}\phi_{12}^{(j)}\left[\sum_{i=0}^{\infty}\phi_{22}^{2i}-\phi_{22}^{2j}\right]&=&\phi_{12}^2\sum_{j=1}^{\infty}\phi_{22}^{2j-2}\left[\sum_{i=0}^{\infty}\phi_{22}^{2i}-\phi_{22}^{2j}\right]=\phi_{12}^2\left[\frac{1}{(1-\phi_{22}^2)^2}-\frac{\phi_{22}^2}{1-\phi_{22}^4}\right].
\\
\sum_{j=0}^{\infty}\phi_{22}^{j}\phi_{12}^{(j)}\left[\sum_{i=0}^{\infty}\phi_{22}^i\phi_{12}^{(i)}-\phi_{22}^j\phi_{12}^{(j)}\right]&=&\phi_{12}^2\phi_{22}^{-2}\sum_{j=1}^{\infty}\phi_{22}^{2j}\left[\sum_{i=1}^{\infty}\phi_{22}^{2i}-\phi_{22}^{2j}\right]=\phi_{12}^2\left[\frac{\phi_{22}^2}{(1-\phi_{22}^2)^2}-\frac{\phi_{22}^2}{1-\phi_{22}^4}\right].
\end{eqnarray*}
On the other hand, for $h>0$ we have
\begin{eqnarray*}
\sum_{j=0}^{\infty}\phi_{12}^{(j)}\phi_{12}^{(j+h)}\phi_{22}^{2h}\left[\sum_{i=0}^{\infty}\phi_{22}^{2i+h}-\phi_{22}^{2j+h}\right]&=&\phi_{12}^2\phi_{22}^{2h}\sum_{j=1}^{\infty}
\phi_{22}^{2j-2}\left[\sum_{i=0}^{\infty}\phi_{22}^{2i}-\phi_{22}^{2j}\right]\\
&=&\phi_{12}^2\phi_{22}^{2h}\left[\frac{1}{(1-\phi_{22}^2)^2}-\frac{\phi_{22}^2}{1-\phi_{22}^4}\right].
\\
\sum_{j=0}^{\infty}\phi_{22}^{j+h}\phi_{12}^{(j)}\left[\sum_{i=0}^{\infty}\phi_{22}^i\phi_{12}^{(i+h)}-\phi_{22}^j\phi_{12}^{(j+h)}\right]&=&\phi_{12}^2\phi_{22}^{2h}\sum_{j=1}^{\infty}\phi_{22}^{2j-1}\left[\sum_{i=0}^{\infty}\phi_{22}^{2i-1}-\phi_{22}^{2j-1}\right]\\
&=&\phi_{12}^2\phi_{22}^{2h}\left[\frac{1}{(1-\phi_{22}^2)^2}-\frac{\phi_{22}^2}{1-\phi_{22}^4}\right].
\\
\sum_{j=0}^{\infty}\phi_{22}^{j+h}\phi_{12}^{(j)}\left[\sum_{m=-h}^{\infty}\phi_{22}^m\phi_{12}^{(m+h)}-\phi_{22}^j\phi_{12}^{(j+h)}\right]&=&\phi_{12}^2\sum_{j=1}^{\infty}\phi_{22}^{2j+h-1}\left[\sum_{m=-h+1}^{\infty}\phi_{22}^{2m+h-1}-\phi_{22}^{2j+h-1}\right]\\&=&\phi_{12}^2\phi_{22}^{2h}\left[\frac{\phi_{22}^{-2h}\phi_{22}^2}{(1-\phi_{22}^2)^2}-\frac{\phi_{22}^2}{1-\phi_{22}^4}\right].
\end{eqnarray*}
Finally, assuming $\phi_{11}=0$ we obtain the following formula for the autocovariance function of the time series $\{Y(t)\}$
\begin{itemize}
    \item If $h=0$\\
\begin{eqnarray}\label{eqn:acvf:case3_h0}
\text{ACVF}_Y(h)&=&Var(Y(t))=\frac{\kappa_Z\phi_{12}^{2}\phi_{22}^2}{1-\phi_{22}^4}+\sigma_{Z,2}^4\phi_{12}^2\left[\frac{1+2\phi_{22}^2}{(1-\phi_{22}^2)^2}-\frac{3\phi_{22}^2}{1-\phi_{22}^4}\right]+\frac{\sigma_{Z,1}^2\sigma_{Z,2}^2}{1-\phi_{22}^2}-\frac{\sigma_{Z,2}^4\phi_{12}^2\phi_{22}^2}{(1-\phi_{22}^2)^2}\nonumber \\
&=&\phi_{12}^{2}\left[\frac{\phi_{22}^2(\kappa_Z-3\sigma_{Z,2}^4)}{1-\phi_{22}^4}+\frac{(1+\phi_{22}^2)\sigma_{Z,2}^4}{(1-\phi_{22}^2)^2}\right]+\frac{\sigma_{Z,1}^2\sigma_{Z,2}^2}{1-\phi_{22}^2}.
\end{eqnarray}
\item If $h>0$\\
\begin{eqnarray}\label{eqn:acvf:case3_h}
\text{ACVF}_Y(h)&=&\frac{\kappa_Z\phi_{12}^{2}\phi_{22}^{2h}\phi_{22}^2}{1-\phi_{22}^4}+\sigma_{Z,2}^4\phi_{12}^2\phi_{22}^{2h}\left[\frac{2+\phi_{22}^{-2h}\phi_{22}^2}{(1-\phi_{22}^2)^2}-\frac{3\phi_{22}^2}{1-\phi_{22}^4}\right]-\frac{\sigma_{Z,2}^4\phi_{12}^2\phi_{22}^2}{(1-\phi_{22}^2)^2}\nonumber \\
&=&\phi_{12}^{2}\phi_{22}^{2h}\left[\frac{\phi_{22}^2(\kappa_Z-3\sigma_{Z,2}^4)}{1-\phi_{22}^4}+\frac{2\sigma_{Z,2}^4}{(1-\phi_{22}^2)^2}\right].
\end{eqnarray}
\end{itemize}

\section{Simulation study} \label{sec:simulations}
To illustrate properties of the product of the VAR(1) model components, we conduct a simulation study. To this end, we simulate bi-dimensional trajectories of the VAR(1) model $\{\textbf{X}(t)\}$ with a bivariate Gaussian or bivariate Student's t  distribution (see Appendix for a detailed description) for the residual vectors. The shape parameter of the Student's t distribution is set to $\eta=5$, so it has much heavier tail than the Gaussian one. For simplicity, we assume that the scale parameters of the Gaussian distribution for both coordinates are equal, i.e. $\sigma_{Z,1}=\sigma_{Z,2}$. Moreover, we assume that $\sigma_{Z,i}=\sqrt{\frac{\eta}{\eta-2}}, i=1,2$, which is equal to the standard deviation of the marginal Student's t distributions. Thus, the variances of both distributions of the residual vectors are equal.  

We analyze the trajectories as well as the autocovariance functions of the product of the simulated VAR(1) vectors defined in Eq. (\ref{for3}) separately for Cases 1-3. We also derive the $5\%$ and $95\%$ confidence bounds (CB) for the autocovariance function using Monte Carlo calculations of the empirical ACVF with 1000 repetitions and two cases of the trajectories length, namely $n=100$ or $n=1000$. Note that the autocovariance at lag $h=0$ is equal to the variance of the time series, ACVF$_Y(0)=\text{Var}(Y(t))$ for each $t$. 

It is worth mentioning that in the case when the residual series has bivariate Gaussian distribution, then $X_1(t)$ and $X_2(t)$ for each $t\in \mathbb{Z}$ have one-dimensional Gaussian distributions and  $Y(t)$ has a variance-gamma distribution  with appropriate parameters \citep[see][]{general_normal}. A detailed analysis related to the distribution of a product Gaussian random variables was presented, for instance, by \cite{nasza_produkt}. In case when $(Z_1,Z_2)$   have the bivariate Student's t distribution, the components of the VAR(1) model $X_1(t)$ and $X_2(t)$ are not Student's t distributed. This is related to the fact that the linear combination of Student's t distributed random variables (with different weights) is not Student's t distributed.

\subsection{Case 1}
In this case the time series $\{X_1(t)\}$ and $\{X_2(t)\}$ are independent. Hence, $\phi_{12}=\phi_{21}=0$ and the residual vectors components $\{Z_1(t)\}$ and $\{Z_2(t)\}$ for each $t\in \mathbb{Z}$ have independent marginal distributions. In the Gaussian case it is simply equivalent to $\rho_Z=0$ in Eq. (\ref{pdf_gaussian_baza}). However, in the Student's t  case putting $\rho_Z=0$ in (\ref{student1}) does not lead to the independence of the marginal distributions. Hence, the components of the residual vector  are simulated as independent one-dimensional Student's t random variables, see Eq. (\ref{student2}) for the PDF formula.

\begin{figure}[h]
    \centering
    \includegraphics[scale=1]{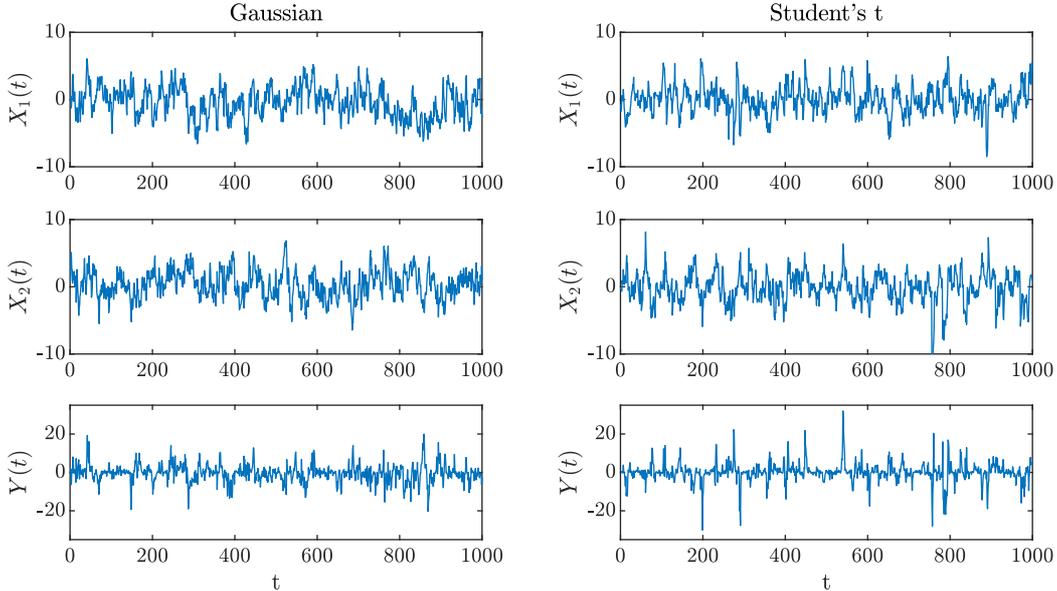}
    \caption{Sample trajectories of the VAR(1) model components and their product for the Gaussian (left panels) and Student's t distribution (right panels). The parameters correspond to Case 1, i.e. $\phi_{11}=0.8$, $\phi_{22}=0.8$, $\phi_{12}=\phi_{21}=0$ and the residual vectors $Z_i(t), i=1,2$ are independent with $\eta=5$ for the Student's t distribution and $\sigma_{Z,1}^2=\sigma_{Z,2}^2=\frac{\eta}{\eta-2}$ for the Gaussian one.}
    \label{fig:case1_same_sgn_traj}
\end{figure}

\begin{figure}[h]
    \centering
    \includegraphics[scale=1]{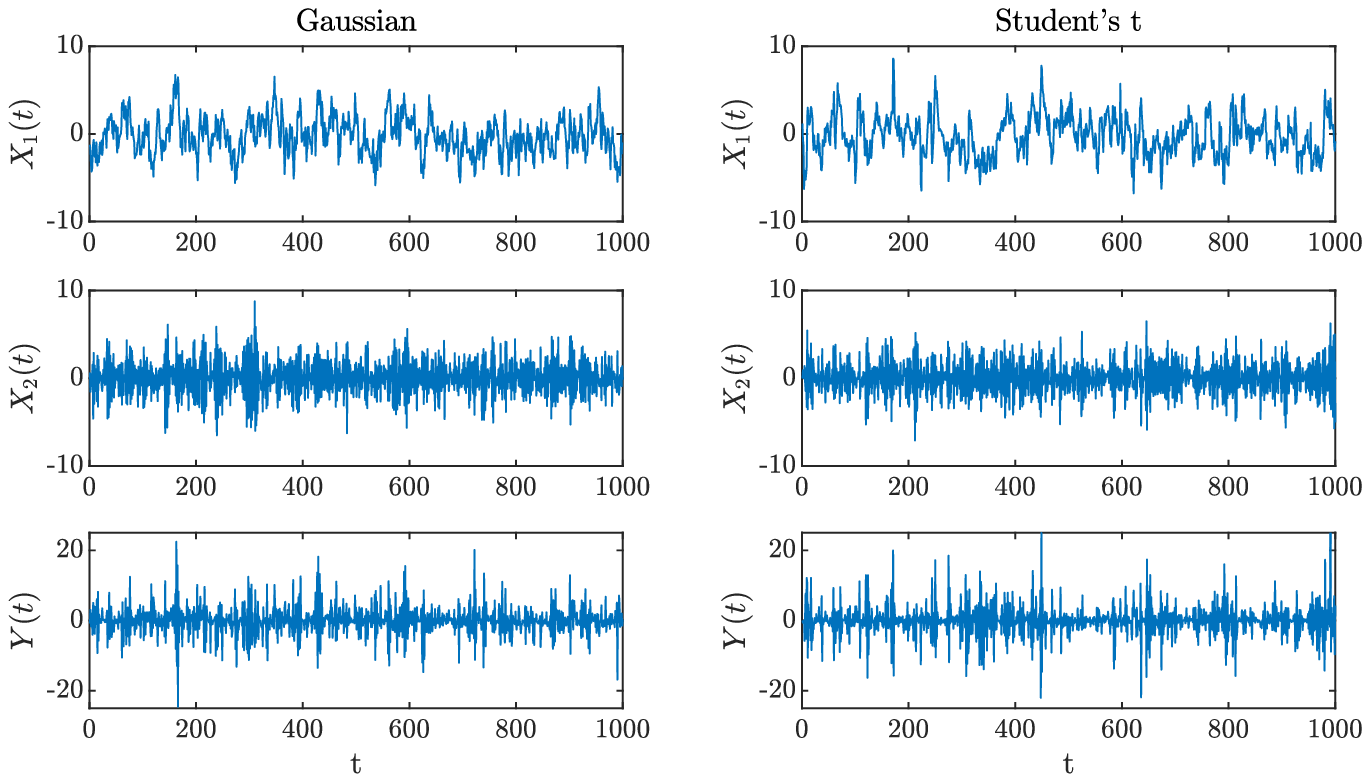}
        \caption{Sample trajectories of the VAR(1) model components and their product for the Gaussian (left panels) and Student's t distribution (right panels). The parameters correspond to Case 1, i.e. $\phi_{11}=0.8$, $\phi_{22}=-0.8$, $\phi_{12}=\phi_{21}=0$ and the residual vectors $Z_i(t), i=1,2$ are independent with $\eta=5$ for the Student's t distribution and $\sigma_{Z,1}^2=\sigma_{Z,2}^2=\frac{\eta}{\eta-2}$ for the Gaussian one.}
    \label{fig:case1_diff_sgn_traj}
\end{figure}

Since the shape of the ACVF of the product time series depends strongly on the sign of $\phi_{11}\phi_{22}$ (see formula (\ref{ex_var_case2})), we use two parameter sets: either $\phi_{11}=0.8$ and $\phi_{22}=0.8$ or $\phi_{11}=0.8$ and $\phi_{22}=-0.8$. The sample trajectories of the model components $\{X_1(t)\}$ and $\{X_2(t)\}$ as well as their product $\{Y(t)\}$ are plotted in Figures \ref{fig:case1_same_sgn_traj} and \ref{fig:case1_diff_sgn_traj} for  $\phi_{22}=0.8$ and  $\phi_{22}=-0.8$, respectively. Comparing both figures a clear difference can be observed in the behaviour of trajectories corresponding to $\{X_2(t)\}$ as well as $\{Y(t)\}$.  Setting $\phi_{11}\phi_{22}<0$ (see Figure \ref{fig:case1_same_sgn_traj}) yields antipersistent behaviour of both time series. Moreover, the distribution of the product time series $\{Y(t)\}$ has heavier tails than for $\{X_1(t)\}$ and $\{X_2(t)\}$. This effect is more pronounced for the Student's t distribution of the residuals. 


\begin{figure}[h]
    \centering
    \includegraphics[scale=0.9]{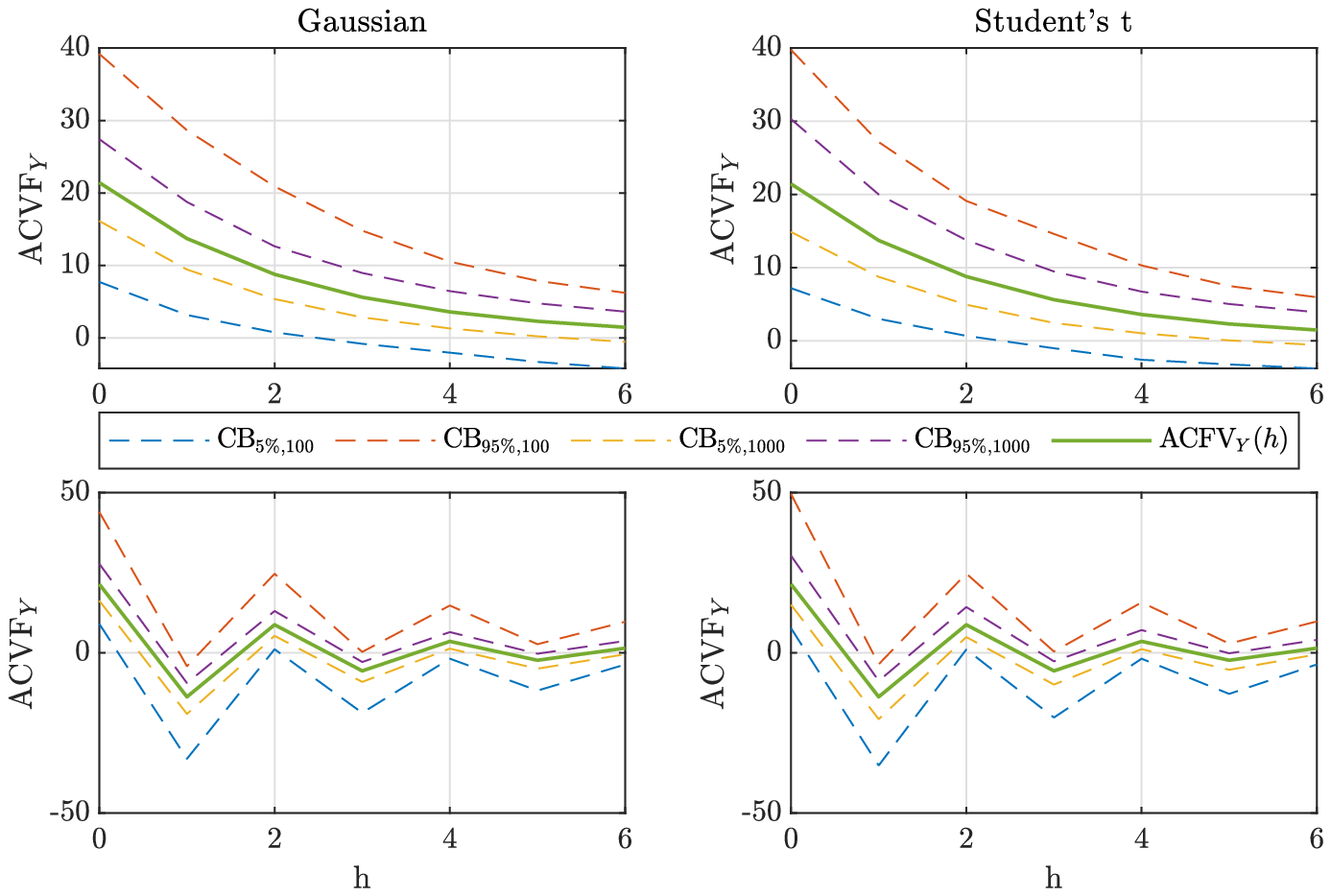}
    \caption{Autocovariance function of the VAR(1) components product ACVF$_Y(h)$ (solid line) and the corresponding confidence bounds $CB_{q,n}$ (dashed lines) for the Gaussian (left panels) and Student's t distribution (right panels). The confidence bounds were obtained using Monte Carlo simulations with 1000 repetitions. The sample length were set to $n=100$ or $n=1000$. The parameters correspond to Case 1,  i.e. $\phi_{11}=0.8$, $\phi_{22}=0.8$, $\phi_{12}=\phi_{21}=0$ (top panels; see Fig. \ref{fig:case1_same_sgn_traj} for the corresponding sample trajectories) or $\phi_{11}=0.8$, $\phi_{22}=-0.8$, $\phi_{12}=\phi_{21}=0$ (bottom panels; see Fig. \ref{fig:case1_diff_sgn_traj} for the corresponding sample trajectories)
    and the residual vectors $Z_i(t), i=1,2$ are independent with $\eta=5$ for the Student's t distribution and $\sigma_{Z,1}^2=\sigma_{Z,2}^2=\frac{\eta}{\eta-2}$ for the Gaussian one.}
    \label{fig:case1_ACVF}
\end{figure}

In Figure \ref{fig:case1_ACVF} we plot ACVF$_Y(h)$ for both sets of parameters and both distributions. The shape of autocovariance function strongly depends on the sign of $\phi_{11}\phi_{22}$. It has a clear power decay if $\phi_{11}\phi_{22}>0$ and antipersistent convergence to zero if $\phi_{11}\phi_{22}<0$. The values of ACVF$_Y(h)$ are the same for both distributions, which follows directly from formula (\ref{ex_var_case2}) and the fact that variances for both distributions are equal. However, a difference between the distributions can be observed in the widths of the intervals between CBs, i.e. the confidence intervals (CIs). Due to the heavier tails for the Student's t distribution wider CIs are obtained in this case. The difference is more apparent if $n=100$, showing that the convergence of the empirical autocovariance to its theoretical value is slower than in the case of the Gaussian distribution.   

\subsection{Case 2}
In this case, the time series $\{X_1(t)\}$ and $\{X_2(t)\}$ are dependent only through the residual vector. Hence, in the Gaussian case we set $\rho_Z\neq0$, while in the Student's t case the bivariate specification (\ref{student1}) need to be used, yielding dependence between $Z_1$ and $Z_2$  for any $\rho_Z\in (-1,1)$. We start with a comparison of the behaviour of the trajectories and autocovariance functions with $\rho_Z=0.8$ in both cases. Note that the sign of $\rho_Z$ does not influence the value of the autocovariance function, what follows directly from formula (\ref{acvfy_final_case3}). On the other hand, similarly as in Case 1, the sign of $\phi_{11}\phi_{22}$ is an important factor for the ACVF behaviour. Hence, we use two parameter sets: either $\phi_{11}=0.8$ and $\phi_{22}=0.8$ or $\phi_{11}=0.8$ and $\phi_{22}=-0.8$. 

\begin{figure}[h]
    \centering
    \includegraphics[scale=1]{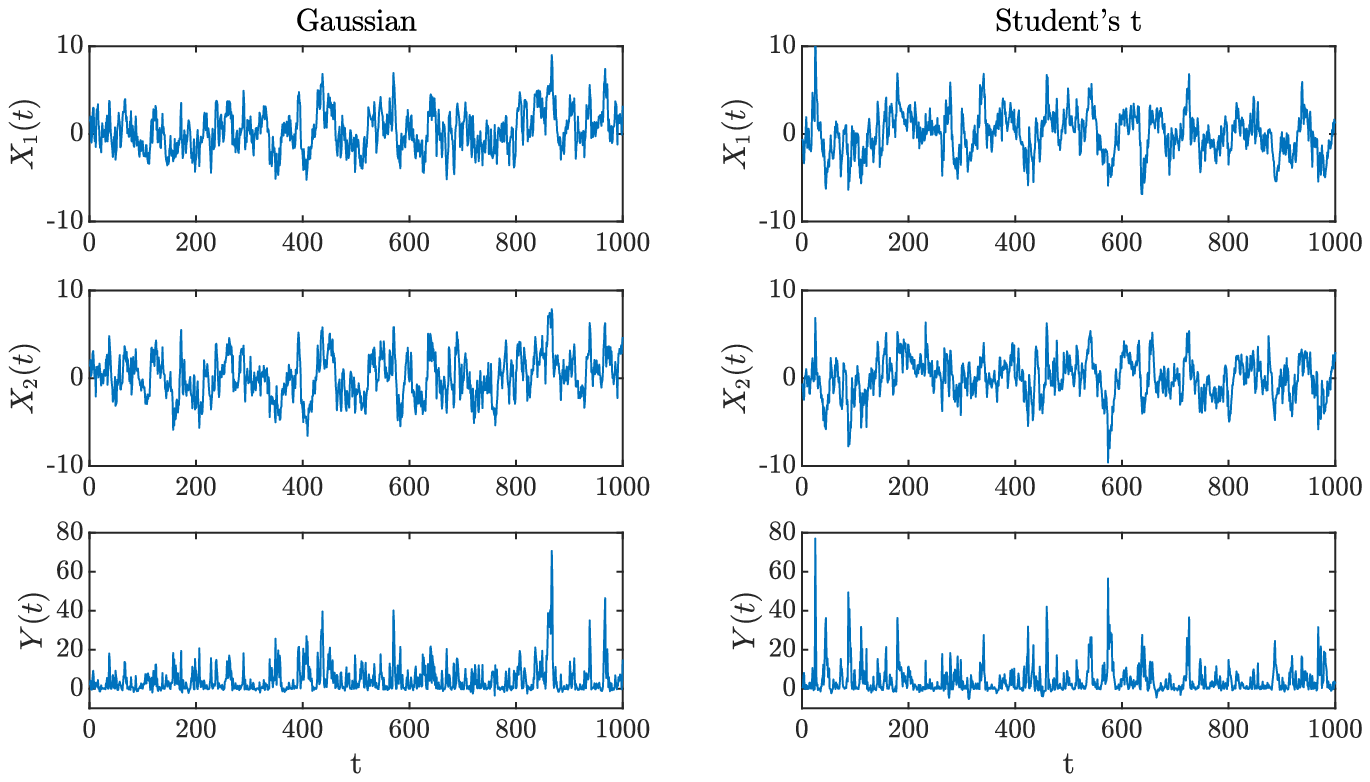}
    \caption{Sample trajectories of the VAR(1) model components and their product for the Gaussian (left panels) and Student's t distribution (right panels). The parameters correspond to Case 2, i.e. $\phi_{11}=0.8$, $\phi_{22}=0.8$, $\phi_{12}=\phi_{21}=0$ and the residual vectors $Z_i(t), i=1,2$ are correlated with $\rho_Z=0.8$. $\eta=5$ for the Student's t distribution and $\sigma_{Z,1}^2=\sigma_{Z,2}^2=\frac{\eta}{\eta-2}$ for the Gaussian one.}
    \label{fig:case2_same_sgn_traj}
\end{figure}

\begin{figure}[h]
    \centering
    \includegraphics[scale=1]{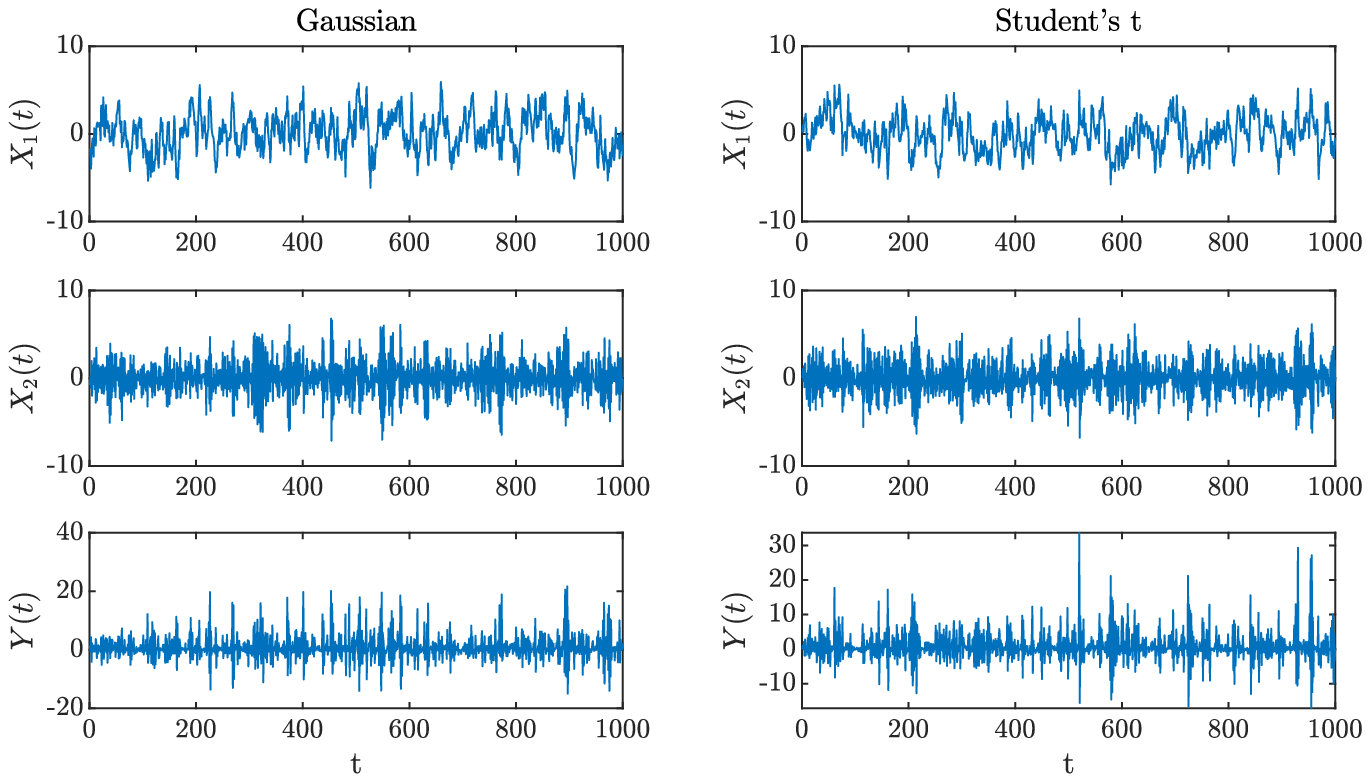}
    \caption{Sample trajectories of the VAR(1) model components and their product for the Gaussian (left panels) and Student's t distribution (right panels). The parameters correspond to Case 2, i.e. $\phi_{11}=0.8$, $\phi_{22}=-0.8$, $\phi_{12}=\phi_{21}=0$ and the residual vectors $Z_i(t), i=1,2$ are correlated with $\rho_Z=0.8$. $\eta=5$ for the Student's t distribution and $\sigma_{Z,1}^2=\sigma_{Z,2}^2=\frac{\eta}{\eta-2}$ for the Gaussian one.}
    \label{fig:case2_diff_sgn_traj}
\end{figure}

The sample trajectories corresponding to the time series $\{X_1(t)\}$, $\{X_2(t)\}$ and their product $\{Y(t)\}$ are plotted in Figures \ref{fig:case2_same_sgn_traj} and \ref{fig:case2_diff_sgn_traj}, respectively. Similarly as in Case 1, we can observe heavier tails for the Student's t distribution and a clear antipersistnecy if $\phi_{11}\phi_{22}<0$. An interesting feature can be observed in Figure \ref{fig:case2_same_sgn_traj}, i.e. when $\phi_{11}\phi_{22}>0$. The strong dependence between the residuals makes the product distribution highly non-symmetric, even though the individual components do not exhibit this feature.

\begin{figure}[h]
    \centering
    \includegraphics[scale=0.9]{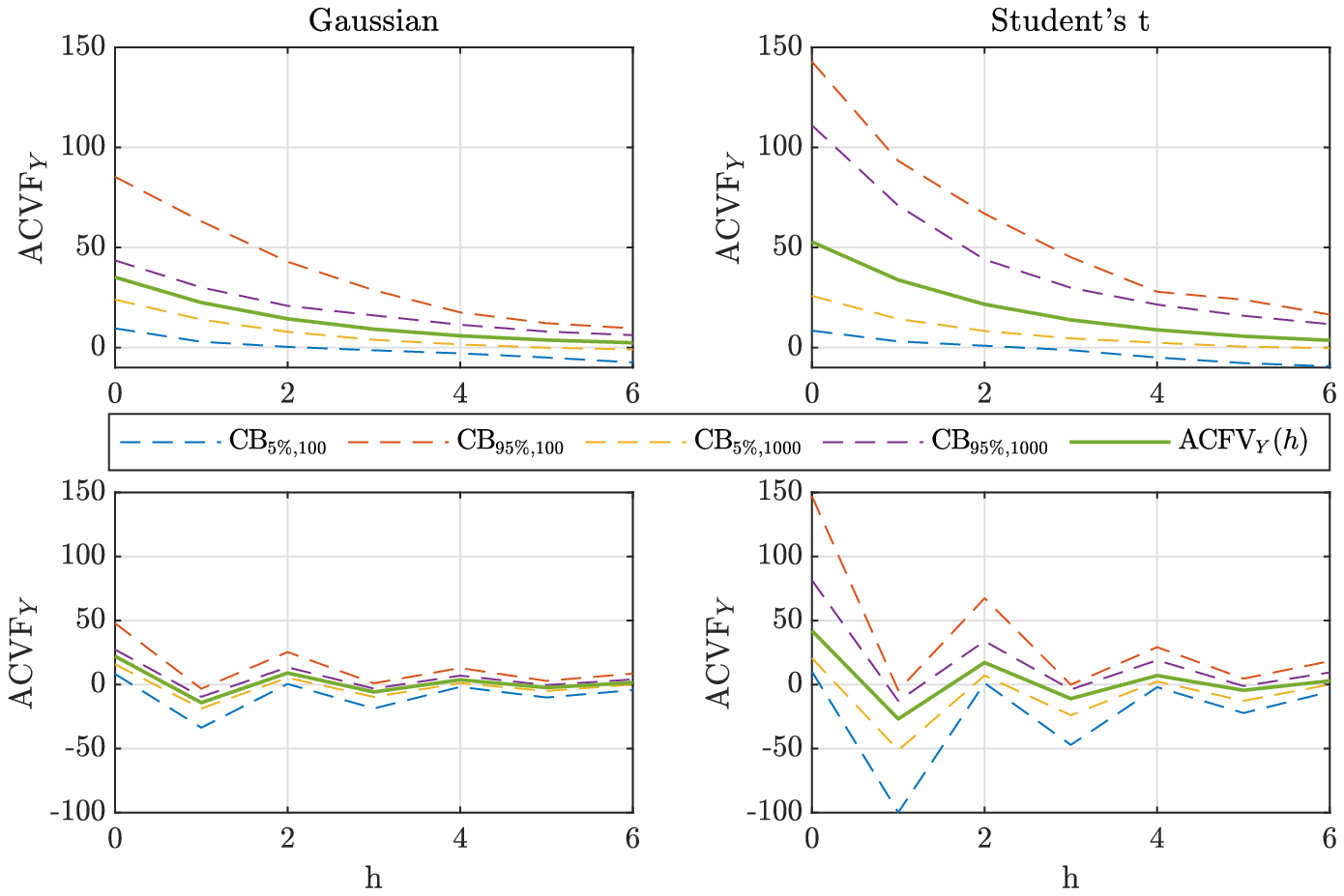}
    \caption{Autocovariance function of the VAR(1) components product ACVF$_Y(h)$ (solid line) and the corresponding confidence bounds CB$_{q,n}$ (dashed lines) for the Gaussian (left panels) and Student's t distribution (right panels). The confidence bounds were obtained using Monte Carlo simulations with 1000 repetitions. The sample length were set to $n=100$ or $n=1000$. The parameters correspond to Case 2,  i.e. $\phi_{11}=0.8$, $\phi_{22}=0.8$, $\phi_{12}=\phi_{21}=0$ (top panels; see Fig. \ref{fig:case2_same_sgn_traj} for the corresponding sample trajectories) or $\phi_{11}=0.8$, $\phi_{22}=-0.8$, $\phi_{12}=\phi_{21}=0$ (bottom panels; see Fig. \ref{fig:case2_diff_sgn_traj} for the corresponding sample trajectories)
    and the residual vectors $Z_i(t), i=1,2$ are correlated with $\rho_Z=0.8$. $\eta=5$ for the Student's t distribution and $\sigma_{Z,1}^2=\sigma_{Z,2}^2=\frac{\eta}{\eta-2}$ for the Gaussian one.}
    \label{fig:case2_ACVF}
\end{figure}

The empirical autocovariance function of the product time series for both parameter sets is plotted in Figure \ref{fig:case2_ACVF}. Again, similarly as in Case 1, we can observe a power decay if $\phi_{11}\phi_{22}>0$ and antipersistency if $\phi_{11}\phi_{22}<0$. However, now the values of ACVF$_Y(h)$ are different for the Gaussian and Student's t distributions. It is a consequence of different values of $m_Z$ in Eq. (\ref{acvfy_final_case3}). For the Gaussian distribution we have the following \citep{nasza_produkt}  
\begin{equation*}
    m_Z=\mathbb{E}\left[Z_1^2Z_2^2\right]=\sigma_{Z,1}^2\sigma_{Z,2}^2+2\rho_Z^2\sigma_{Z,1}^2\sigma_{Z,2}^2.
\end{equation*}
Hence, the autocovariance function simplifies to
\begin{eqnarray*}
\text{ACVF}_Y(h)
=\left(\phi_{11}\phi_{22}\right)^h\left[\frac{\sigma_{Z,1}^2\sigma_{Z,2}^2}{(1-\phi_{11}^2)(1-\phi_{22}^2)}+\frac{\rho_Z^2\sigma_{Z,1}^2\sigma_{Z,2}^2}{\left(1-\phi_{11}\phi_{22}\right)^2}\right].
\end{eqnarray*}
This is not the case for the Student's t distribution, for which in general $m_Z\neq\sigma_{Z,1}^2\sigma_{Z,2}^2+2\rho_Z^2\sigma_{Z,1}^2\sigma_{Z,2}^2$, so the first component in the ACVF$_Y(h)$ (see formula  (\ref{acvfy_final_case3})) is non-zero. 

The difference between the distributions is also clearly visible in the widths of the confidence intervals. For the Student's t distribution much longer trajectories are needed to obtain a good convergence of the empirical autocovariance to its theoretical values.   

\begin{figure}[h]
    \centering
    \includegraphics[scale=0.9]{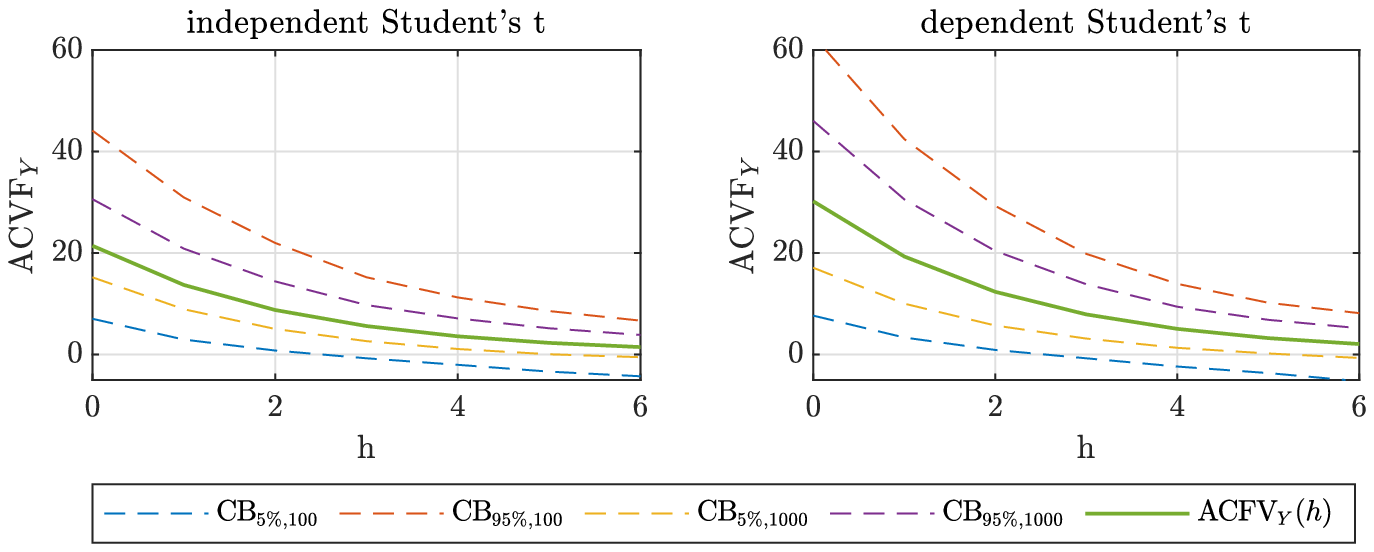}
    \vspace{-3.5cm}
    \caption{Autocovariance function of the VAR(1) components product ACVF$_Y(h)$ (solid line) and the corresponding confidence bounds CB$_{q,n}$ (dashed lines) for the bivariate Student's t distribution (right panel) with $\rho_Z=0$ and independent Student's distributed residual vectors $Z_i(t), i=1,2$ (left panel). The confidence bounds were obtained using Monte Carlo simulations with 1000 repetitions. The sample length were set to $n=100$ or $n=1000$.  The model parameters are equal $\phi_{11}=0.8$, $\phi_{22}=0.8$, $\phi_{12}=\phi_{21}=0$,  $\eta=5$.}
    \label{fig:case1_case2_ACVF}
\end{figure}

Since for the Student's t distribution $\rho_Z=0$ is not equivalent to independence, we also compare the ACVF$_Y(h)$ values obtained with $\rho_Z=0$ in Case 2 (i.e. with the bivariate Student's distribution, defined in Eq. (\ref{student1})) with the ones obtained in Case 1 (i.e. with independent one-dimensional Student's t residuals). Recall that for the Gaussian case putting $\rho_Z=0$ leads to independence, hence it directly corresponds to Case 1. The ACFV$_Y$ functions for both Student's t specifications are plotted in Figure \ref{fig:case1_case2_ACVF}. Indeed, we can observe that the dependence between $Z_1$ and $Z_2$ changes the values of ACVF$_Y(h)$, even if $\rho_Z=0$. It is a consequence of the $m_Z$ value in formula (\ref{acvfy_final_case3}).

\subsection{Case 3}
 In this case, the random variables  $Z_1(t)$ and $Z_2(t)$ are independent for each $t\in \mathbb{Z}$ and the relation between $\{X_1(t)\}$ and $\{X_2(t)\}$ is driven only by  $\phi_{12}\neq 0$ and $\phi_{22}\neq0$. Hence, we simulate the trajectories using the Gaussian distribution with $\rho_Z=0$ and the independent one-dimensional Student's t distributions. Note that it follows directly from formulas (\ref{eqn:acvf:case3_h0}) and (\ref{eqn:acvf:case3_h}), that, differently than in Case 1 and Case 2, the sign of $\phi_{12}\phi_{22}$ does not influence the autocovariance values. Hence, now we use only one parameter set, namely $\phi_{12}=0.8$, $\phi_{22}=0.8$, $\phi_{11}=\phi_{21}=0$.  
 
\begin{figure}[h]
    \centering
    \includegraphics[scale=1]{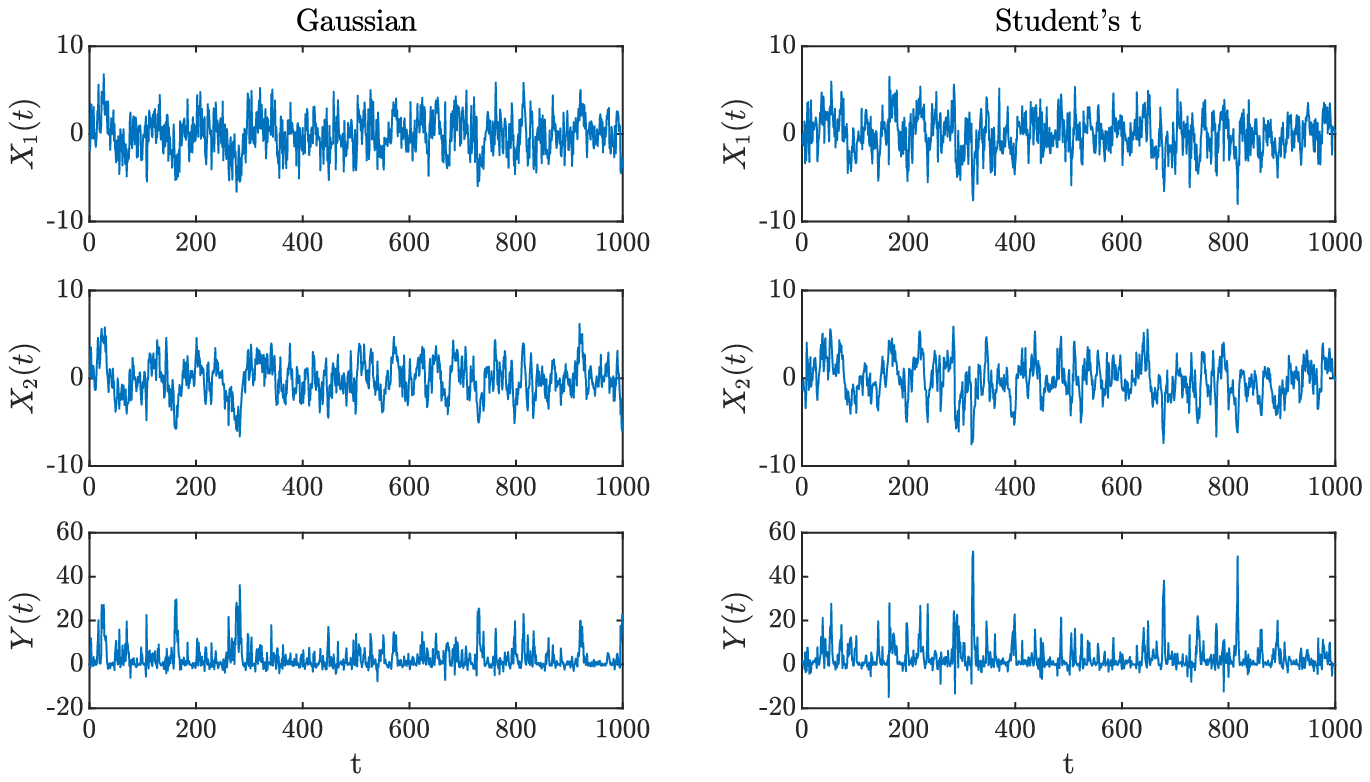}
    \caption{Sample trajectories of the VAR(1) model components and their product for the Gaussian (left panels) and Student's t distribution (right panels). The parameters correspond to Case 3, i.e. $\phi_{12}=0.8$, $\phi_{22}=0.8$, $\phi_{11}=\phi_{21}=0$ and the residual vectors $Z_i(t), i=1,2$ are independent with $\eta=5$ for the Student's t distribution and $\sigma_{Z,1}^2=\sigma_{Z,2}^2=\frac{\eta}{\eta-2}$ for the Gaussian one.}
    \label{fig:case3_traj}
\end{figure}

 The sample trajectories corresponding to time series $\{X_1(t)\}$, $\{X_2(t)\}$ and $\{Y(t)\}$ are plotted in Figure \ref{fig:case3_traj}. Similarly as in Case 2, asymmetry of the product can be observed. However, now it is the effect of dependence through $\phi_{12}$ and $\phi_{22}$ coefficients and not the residuals correlation. 

\begin{figure}[h]
    \centering
    \includegraphics[scale=0.9]{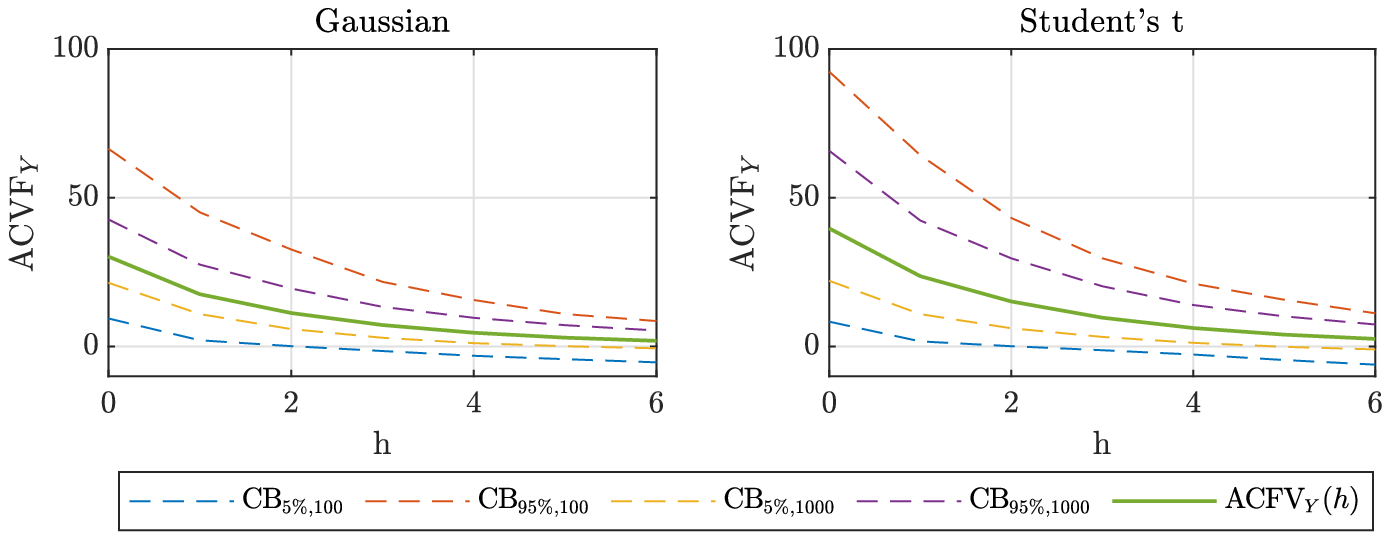}
    \vspace{-3.5cm}
    \caption{Autocovariance function of the VAR(1) components product ACVF$_Y(h)$ (solid line) and the corresponding confidence bounds CB$_{q,n}$ (dashed lines) for the Gaussian (left panel) and Student's t distribution (right panel). The confidence bounds were obtained using Monte Carlo simulations with 1000 repetitions. The sample length were set to $n=100$ or $n=1000$. The parameters correspond to Case 3,  i.e. $\phi_{12}=0.8$, $\phi_{22}=0.8$, $\phi_{11}=\phi_{21}=0$ (see Fig. \ref{fig:case3_traj} for the corresponding sample trajectories) 
    and the residual vectors $Z_i(t), i=1,2$ are independent with $\eta=5$ for the Student's t distribution and $\sigma_{Z,1}^2=\sigma_{Z,2}^2=\frac{\eta}{\eta-2}$ for the Gaussian one.}
    \label{fig:case3_ACVF}
\end{figure}

 The autocovariance functions of the product time series together with the corresponding confidence bounds are plotted in Figure \ref{fig:case3_ACVF}. For both distributions we can observe a power decay of ACVF$_Y(h)$, but the values are different. 
 For the Gaussian distribution we have
 \begin{equation}
     \kappa_Z=\mathbb{E}\left[Z_2^4\right]=3\sigma_{Z,2}^4,
 \end{equation}
 what follows directly from the fact that the kurtosis, i.e. $\mathbb{E}[Z^4]\slash[\mathbb{V}\text{ar}(Z)]^2$, of the Gaussian distribution is equal to 3. This is not the case for the Student's t distribution. Hence, in the Gaussian case, ACVF$_Y(h)$ simplifies to
\begin{equation*}
\text{ACVF}_Y(h)=
\begin{cases}
\phi_{12}^{2}\left[\frac{(1+\phi_{22}^2)\sigma_{Z,2}^4}{(1-\phi_{22}^2)^2}\right]+\frac{\sigma_{Z,1}^2\sigma_{Z,2}^2}{1-\phi_{22}^2}, \quad\text{ if } \quad  h=0,\\
\phi_{12}^{2}\phi_{22}^{2h}\left[\frac{2\sigma_{Z,2}^4}{(1-\phi_{22}^2)^2}\right], \quad \quad \quad \quad \quad \text{ if } \quad  h>0,
\end{cases}
\end{equation*}
while for the Student's t the first component in formulas (\ref{eqn:acvf:case3_h0}) and (\ref{eqn:acvf:case3_h}) is always positive, making the ACVF$_Y(h)$ higher than in the Gaussian case. The confidence intervals are again wider for the Student's t distribution.

\section{Electricity market case study} \label{sec:case_study}
 Electricity prices are known to be autoregressive and highly dependent on the physical demand (or equivalently load), see e.g. \cite{RWeron_energy}. The day-ahead forecasts of the load are usually  published by the transmission system operators (TSO) and can be used for cost or production planning in electricity companies. However, these forecasts are burdened with prediction errors, which might cause a large deviations of the actual energy cost from its predictions. The cost of these errors is equal to the product of the price and the difference between the actual and forecasted load. Hence, the methodology derived in Section \ref{sec:theory} might be useful in such a context. 

\begin{figure}[h] 
    \centering
    \includegraphics{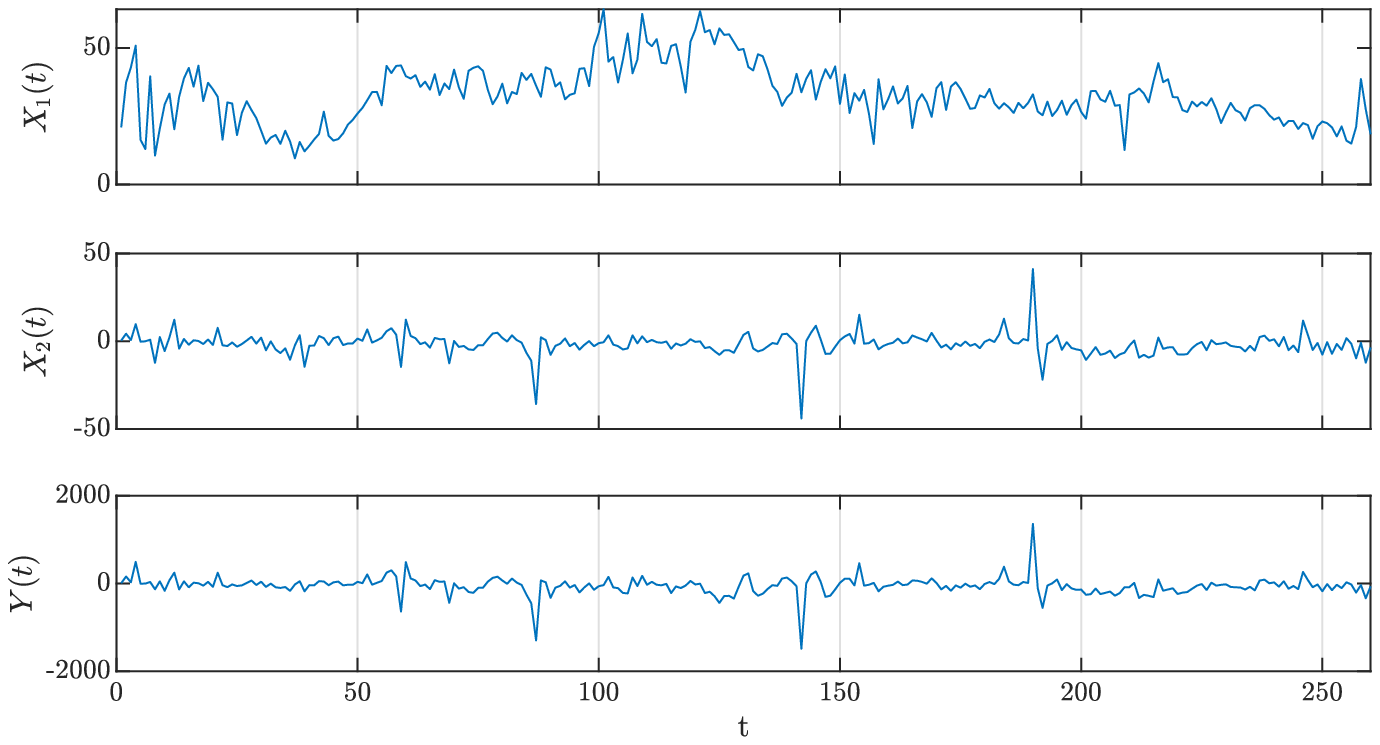}
    \caption{The analyzed Danish electricity market data from the time period 1.1.2016-31.12.2020. Top panel: the weekly means of the day-ahead electricity prices (corresponding to time series $\{X^{\mu}_1(t)\}$). Middle panel: the weekly means of the load TSO forecast errors (corresponding to time series $\{X_2(t)\}$). Bottom panel: the product of weekly day-ahead prices and load forecast errors, ($\{Y(t)\}=\{X^{\mu}_1(t)X_2(t)\}$), i.e the total cost of these errors.}
    \label{fig:case_study_variables}
\end{figure}

We apply the VAR(1) model to electricity day-ahead market data from the danish DK1 zone spanning over time period 1.1.2016-31.12.2021, available from  \cite{ENTSOE}. The time series corresponding to the first variable, denoted as $\{X^{\mu}_1(t)\}$, are the weekly means of the day-ahead electricity prices, while the second variable, corresponding to the time series $\{X_2(t)\}$, are the weekly means of the load forecast errors, i.e. the difference between the load values forecasted by the TSO and the corresponding actual values. The product of these time series $\{Y(t)\}=\{X^{\mu}_1(t)X_2(t)\}$ is the total weekly cost of the load prediction errors. The analyzed time series are plotted in Figure \ref{fig:case_study_variables}.   

\begin{figure}[h] 
    \centering
    \includegraphics{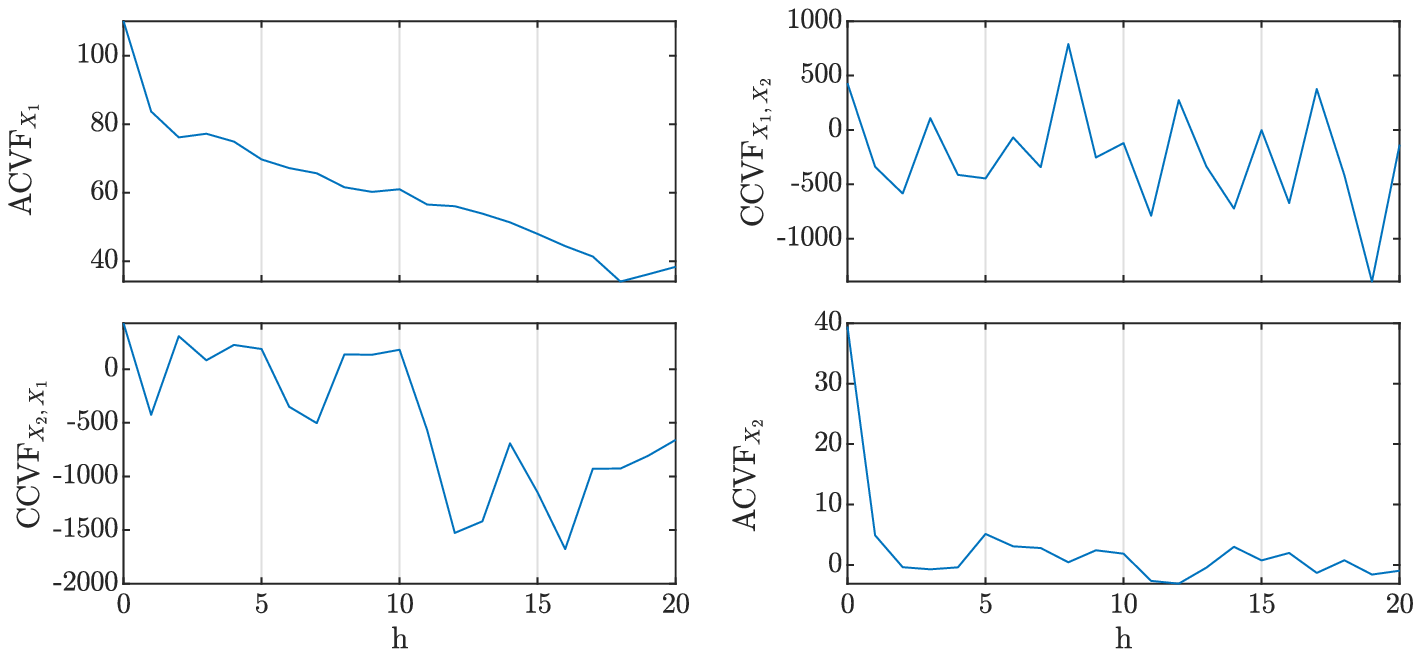}
    \caption{The empirical autocovariance function for time series corresponding to $\{X_1(t)\}$, ACVF$_{X_1}(h)$ (top, left panel) and $\{X_2(t)\}$, ACVF$_{X_2}(h)$) (bottom, right panel), as well as the corresponding empirical cross-covariances CCVF$_{X_1,X_2}(h)$ (top, right panel) and CCVF$_{X_2,X_1}(h)$ (bottom, right panel).}
    \label{fig:case_study_variables_acvfs}
\end{figure}

\begin{figure}[h] 
    \centering
    \includegraphics{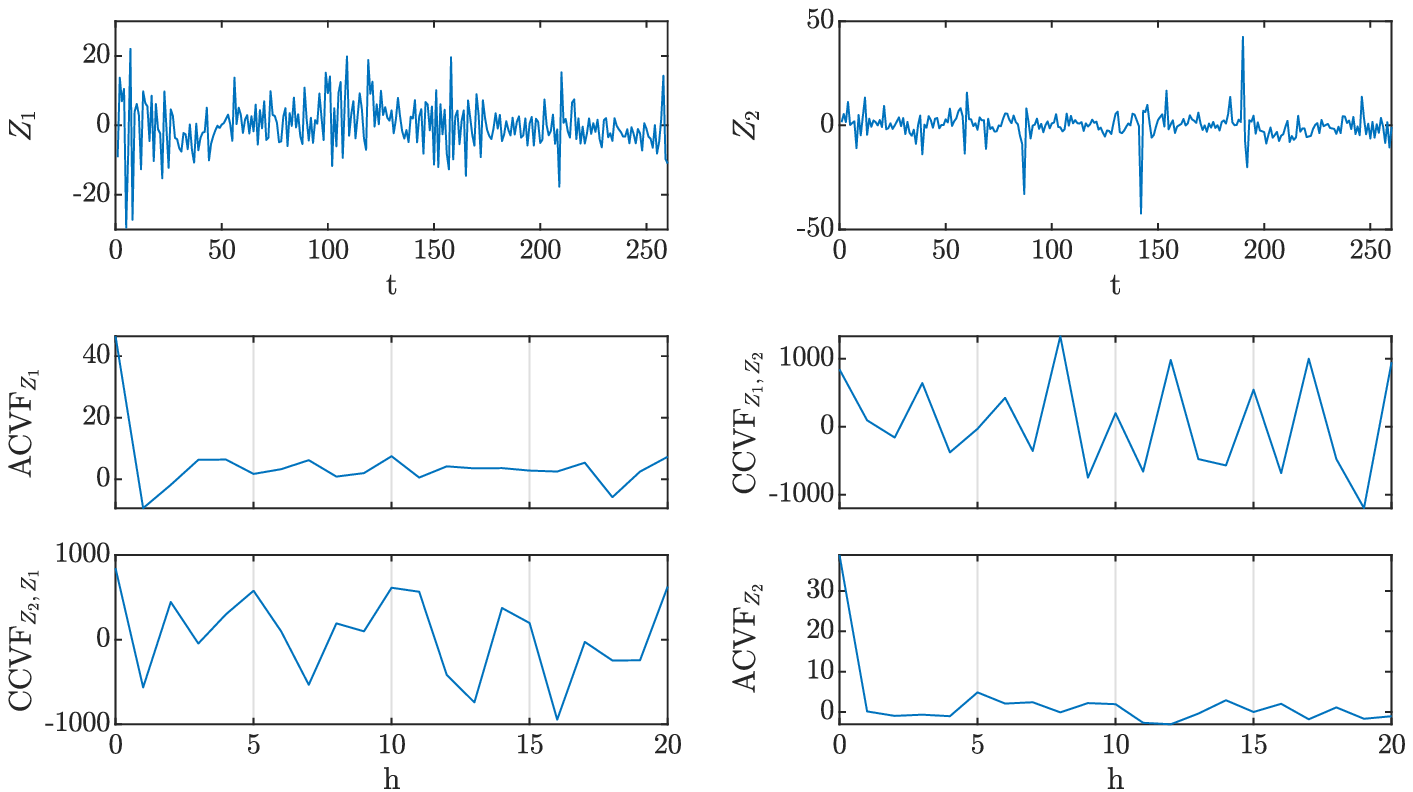}
    \caption{The residual series corresponding to $\{Z_1(t)\}$, $\{Z_2(t)\}$ (top right and left panels, respectively) and the empirical autocovariance function ACVF$_{Z_1}$ (middle, left panel), ACVF$_{Z_2}$ (bottom, right panel), as well as the empirical crosscovariances CCVF$_{Z_1,Z_2}$ (middle, right panel) and CCVF$_{Z_2,Z_1}$ (bottom, right panel).}
    \label{fig:case_study_residuals_acvfs}
\end{figure}

Before applying the VAR(1) model, the electricity prices are demeaned, i.e., for each $t$ we calculate  $X_1(t)=X^{\mu}_1(t)-\mu_X$, where $\mu_X$ is the mean of the prices corresponding to $\{X^{\mu}_1(t)\}$. In Figure \ref{fig:case_study_variables_acvfs} we plot the empirical ACVF obtained for both time series corresponding to $\{X_1(t)\}$ and $\{X_2(t)\}$, as well as the empirical cross-covariances, $\text{CCVF}_{X_1,X_2}(h)=\text{Cov}(X_1(t),X_2(t+h))$, between both components. 
The shapes of the obtained curves indicate a strong autoregressive effect in  $\{X_1(t)\}$, a lower one in  $\{X_2(t)\}$  and no such effects between the components. This observation is confirmed by the estimated matrix $\Phi$ of the coefficients of the VAR(1) model obtained using the Yule-Walker algorithm \citep[see e.g.][]{brockwell2016introduction} 
\begin{equation}\label{case_study_var_coefficients}
    {\Phi}=\left[
    \begin{array}{cccc}
    \phantom{-}0.7639 & -0.0629\\
    -0.0167& \phantom{-}0.1247\\
    \end{array}
    \right].
\end{equation}
Next, we analyze the residuals  obtained from the fitted VAR(1) model. The one-dimensional time series corresponding to $\{Z_1(t)\}$ and $\{Z_2(t)\}$ are plotted in Figure \ref{fig:case_study_residuals_acvfs} together with the corresponding empirical auto- and cross-covariances. The obtained curves are close to 0 and show no time dependence in the residual series. 
The empirical correlation between the trajectories corresponding to $\{Z_1(t)\}$ and $\{Z_2(t)\}$ is equal to $\rho_{Z}=0.0766$ and is not significantly different from 0 according to the $t$ test, \citep[see e.g.][]{tests_book}, as the $p$-value is equal to $0.2183$. The independence of residuals' components  is further confirmed by the $\chi^2$ independence test, \citep[see e.g.][]{tests_book}, which yields the $p$-value of $0.2395$.

\begin{figure}[h] 
    \centering
    \includegraphics{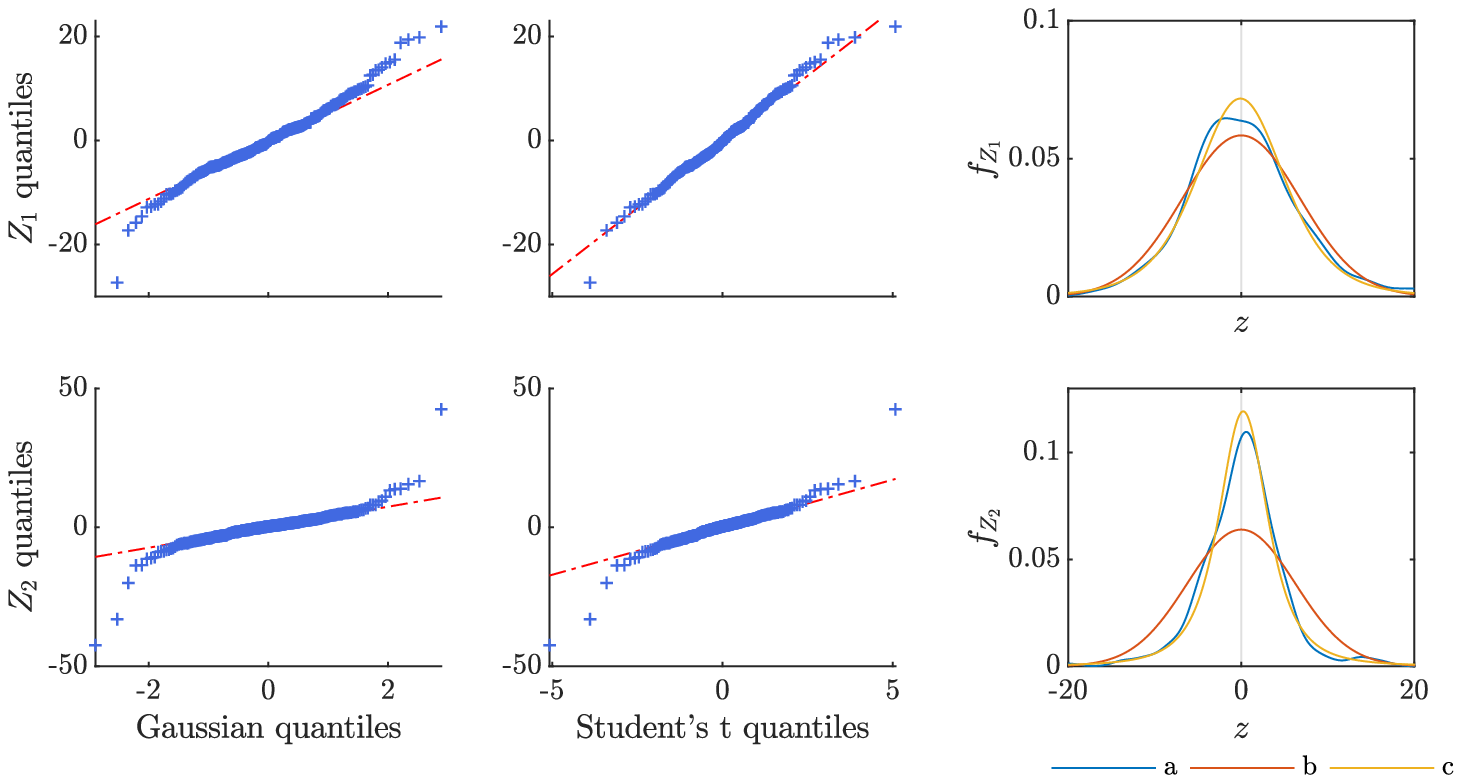}
    \caption{The quantile-quantile plots for the Gaussian (left panels) and Student's t location-scale distribution (middle panels) fitted to time series corresponding to $\{Z_1(t)\}$ (top panels) and $\{Z_2(t)\}$ (bottom panels). In the right panels the empirical PDFs (a, blue colour) corresponding to $\{Z_1(t)\}$, $f_{Z_1}(z)$, and $\{Z_2(t)\}$, $f_{Z_2}(z)$, together with the fitted PDFs for the Gaussian (b, red colour) and Student's t location-scale (c, yellow colour) distributions are plotted.}
    \label{fig:case_study_residuals_qq_and_densities}
\end{figure}

Finally, we fit a distribution to the components of the residual series. Since the independence between time series corresponding to $\{Z_1(t)\}$ and $\{Z_2(t)\}$ for each $t$ cannot be rejected, we analyze them separately as one-dimensional samples. In Figure \ref{fig:case_study_residuals_qq_and_densities} we plot the empirical probability density functions (PDFs) of sample trajectories corresponding to  $\{Z_1(t)\}$ and $\{Z_2(t)\}$ together with the fitted PDFs of the Gaussian and Student's t location-scale distributions (see Appendix for details) and the corresponding quantile-quantile plots. It can be observed that the Student's t distribution yields a good fit to the empirical PDF and there is much improvement over the Gaussian one, especially in the case of load prediction errors (corresponding to $\{Z_2(t)\}$). The  fit is further confirmed by the Kolmogorov-Smirnov goodness-of-fit test, \citep[see e.g.][]{tests_book}. For the trajectory corresponding to the first component, the obtained $p$-values are equal to $0.1063$ or $0.7458$ for the zero mean Gaussian or Student's t location-scale distribution with $\mu=0$, respectively. For the trajectory corresponding to  $\{Z_2(t)\}$ the $p$-values are equal to $0.0012$ and $0.5016$, respectively. Hence, the Student's t distribution with the scale parameter cannot be rejected at any reasonable significance level for both variables, while the Gaussian distribution can be rejected in the case of the time series corresponding to $\{Z_2(t)\}$.      

Summarizing all of the obtained results, we conclude that a good fit is obtained for Case 1 (see Section \ref{special_cases}) of the VAR(1) model with Student's t distributed residuals, i.e. $\phi_{12}=\phi_{21}=0$ and $\{Z_1(t)\}$ and $\{Z_2(t)\}$ are independent. Estimation of the model parameters under Case 1 specification yields: $\phi_{11}=0.7630$, $\phi_{22}=0.1241$, while the Student's t degrees of freedom parameters $\eta_{Z,1} = 4.85$, $\eta_{Z,2} = 2.47$ and the scale parameters are equal to $\lambda_{Z,1} = 5.28$,  $\lambda_{Z,2} = 3.03$.    

Recall that the VAR(1) model was fitted to the demeaned prices, i.e. it was assumed that to $X_1(t)= X^{\mu}_1(t) - \mu_X$, where $\mu_X$ is the mean of the prices $\{X_1^{\mu}(t)\}$. Hence, before analyzing the final cost of the error, i.e.
\begin{equation}
Y(t)=(X_1(t)+\mu_X)X_2(t),
\end{equation}
there is a need to apply the mean-shift also to the first coordinate of the fitted VAR(1) model. In Case 1 straightforward derivations lead to the following formula for the autocovariance of $\{Y(t)\}$ 
\begin{equation}\label{eqn:acvf_mu}
\text{ACVF}_Y(h)=\text{ACVF}_{X_1X_2}(h) + \mu_X^2 \text{ACVF}_{X_2}(h)=
\frac{\sigma_{Z,1}^2\sigma_{Z,2}^2(\phi_{11}\phi_{22})^h}{(1-\phi_{11}^2)(1-\phi_{22}^2)} + \mu_{X_1}^2 \phi_{22}^h \frac{\sigma_{Z,2}^2}{1-\phi_{22}^2 }. 
\end{equation}

\begin{figure}[h] 
    \centering
    \includegraphics{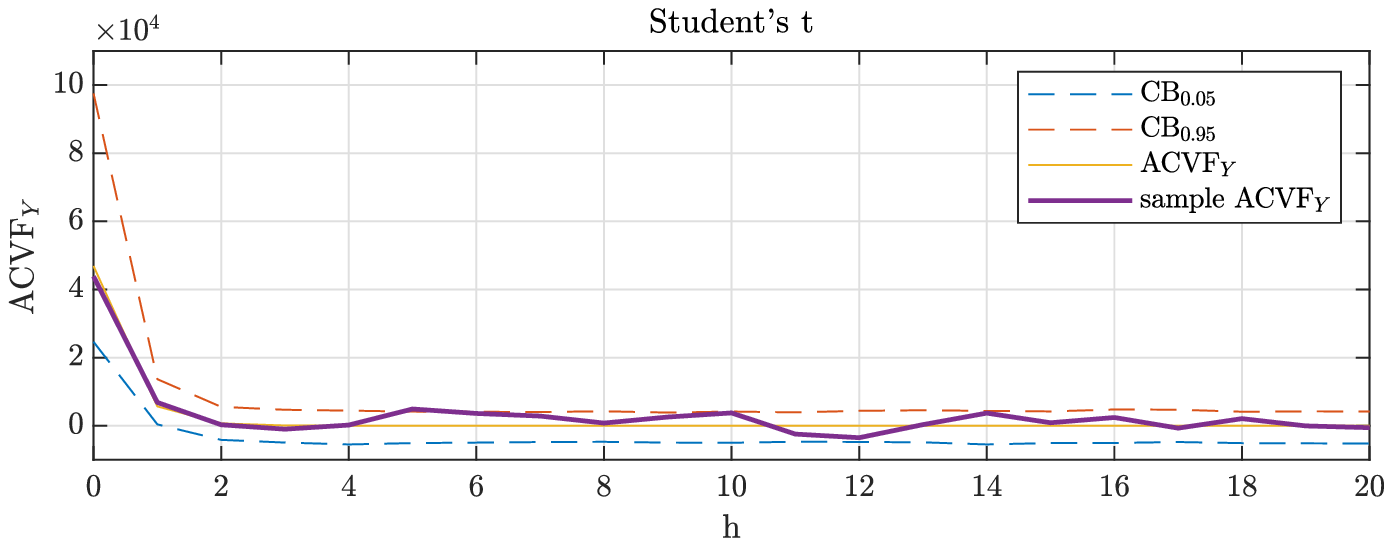}
    \caption{The autocovariance function of the product of prices and load prediction errors, ACVF$_Y(h)$. The empirical autocovariance is plotted with a violet colour, while the theoretical values calculated for the fitted VAR(1) model, see Eq. (\ref{eqn:acvf_mu}) are plotted with a yellow colour. Additionally $5\%$ and $95\%$ CBs obtained from Monte Carlo simulations of the fitted model are plotted with dashed lines.} 
    \label{fig:case_study_product_acvf}
\end{figure}

In Figure \ref{fig:case_study_product_acvf} we plot the derived theoretical autocovariance function for $\{Y(t)\}$, see Eq. (\ref{eqn:acvf_mu}), with the estimated parameters and the fitted Student's t residual distribution, as well as  the empirical autocovariance function calculated for the product of the analyzed data.  
Additionally, we also calculate the $5\%$ and $95\%$ CBs for the autocovariance function of the fitted model using Monte Carlo simulations with 1000 repetitions. As can be observed, the empirical autocovariance curve resembles the shape of the theoretical values and lies within the confidence intervals. It confirms that the fitted VAR(1) model describes well the properties of the product. Hence, the presented approach provides a well fitted model for both the prices and the load prediction errors and at the same time allows for modeling the total cost of the TSO load forecast errors, being the product of both variables. The results might be useful in cost planning for energy companies and help in a proper evaluation of the risk related to the errors of the load/demand predictions.  

\section{Conclusions} \label{sec:conclusions}

In this paper, we have introduced a new times series arising as a product of the bi-dimensional VAR(1) model components and derived formulas for its main characteristics, such as the mean and the autocovariance function. The obtained results describe the time-dependence structure of the product time series and depend on the residual series's distribution only through its parameters. However, the distribution of the product time series can be also derived under the assumption on the form of the residual series's distribution using the probabilistic  properties of the product and the sum of random variables. Specifically, for the Gaussian case analyzed in the simulation study, the resulting distribution of the product time series is the variance gamma. Clearly, the results obtained in this paper can be further generalized for other time series models, especially for the VAR model with higher order or higher dimension.  In the literature, there are also considered VAR models with heavy-tailed multidimensional distribution, e.g. $\alpha-$stable \citep{ola22,grzesiekfloc,ola55}, thus the natural extension of the current study is the analysis of  time series that is a product of two components of such models. In this case, the dependence structure can not be expressed by the means of the autocovariance function  defined for finite-variance models, but by the dependence measures properly defined for the infinite-variance time series, see e.g. \cite{ola11}.      

Since a product variable occurs naturally in many real life processes, we believe that the derived theoretical results can find many practical applications. In the paper, we have conducted a case study based on the data from the Danish electricity market. We have shown that the weekly prices and load prediction errors can be modelled by a VAR(1) model, which also yields a good fit for the product time series, describing in this case the total cost of the load prediction errors. As the load forecasts are a crucial parameter for production and trade planning in electricity companies, the proper evaluation of the risk of their errors is essential for market strategies planing. 

Another possible application of the obtained results is modelling of commodity prices in the currency of the country where a company operates. Commodities prices are fixed at main commodities exchanges where they are quoted usually in USD. Then the price of a commodity in local currency is just the product of the price in USD and the currency exchange rate. On the other hand, the relation between commodity price in USD and local currency exchange rate throughout the time is extremely weighty as it has a significant impact on entity's profitability and can be modelled using VAR model. A recent analysis of such modelling approach for a mining company was conducted by \cite{reso}.    

\section*{Acknowledgments}
J.J. and A.P. acknowledges a support of NCN Sonata Grant No. 2019/35/D/HS4/00369. The work of A.W. was supported by National Center of Science under Opus Grant No. 2020/37/B/HS4/00120 "Market risk model identification and validation using novel statistical, probabilistic, and machine learning tools". 

\section*{Data availability statement}
The electricity market data are available from \cite{ENTSOE}.

\bibliography{mybibliography}

\begin{thebibliography}{54}
\expandafter\ifx\csname natexlab\endcsname\relax\def\natexlab#1{#1}\fi
\providecommand{\url}[1]{\texttt{#1}}
\providecommand{\href}[2]{#2}
\providecommand{\path}[1]{#1}
\providecommand{\DOIprefix}{doi:}
\providecommand{\ArXivprefix}{arXiv:}
\providecommand{\URLprefix}{URL: }
\providecommand{\Pubmedprefix}{pmid:}
\providecommand{\doi}[1]{\href{http://dx.doi.org/#1}{\path{#1}}}
\providecommand{\Pubmed}[1]{\href{pmid:#1}{\path{#1}}}
\providecommand{\bibinfo}[2]{#2}
\ifx\xfnm\relax \def\xfnm[#1]{\unskip,\space#1}\fi
\bibitem[{Adamska et~al.(2021)Adamska, Bielak, Janczura and
  Wy{\l}oma{\'n}ska}]{nasza_produkt}
\bibinfo{author}{Adamska, J.}, \bibinfo{author}{Bielak, {\L}.},
  \bibinfo{author}{Janczura, J.}, \bibinfo{author}{Wy{\l}oma{\'n}ska, A.},
  \bibinfo{year}{2021}.
\newblock \bibinfo{title}{On the distribution of the product of two continuous
  random variables with an application to electricity market transactions.
  {F}inite and infinite-variance case}.
\newblock \bibinfo{journal}{arxiv} \bibinfo{volume}{arXiv:2111.13487}.
\bibitem[{Aroian et~al.(1978)Aroian, Taneja and Cornwell}]{general_normal}
\bibinfo{author}{Aroian, L.A.}, \bibinfo{author}{Taneja, S.V.},
  \bibinfo{author}{Cornwell, L.W.}, \bibinfo{year}{1978}.
\newblock \bibinfo{title}{Mathematical forms of the distribution of the product
  of two normal variables}.
\newblock \bibinfo{journal}{Communications in Statistics - Theory and Methods}
  \bibinfo{volume}{7}, \bibinfo{pages}{165--172}.
\bibitem[{Bhargav et~al.(2018)Bhargav, da~Silva, Chun, Leonardo, Cotton and
  Yacoub}]{appl8}
\bibinfo{author}{Bhargav, N.}, \bibinfo{author}{da~Silva, C.R.N.},
  \bibinfo{author}{Chun, Y.J.}, \bibinfo{author}{Leonardo, E.J.},
  \bibinfo{author}{Cotton, S.L.}, \bibinfo{author}{Yacoub, M.D.},
  \bibinfo{year}{2018}.
\newblock \bibinfo{title}{On the product of two $\kappa$ – $\mu$ random
  variables and its application to double and composite fading channels}.
\newblock \bibinfo{journal}{IEEE Transactions on Wireless Communications}
  \bibinfo{volume}{17}, \bibinfo{pages}{2457--2470}.
\bibitem[{Bhargava and Khatri(1981)}]{beta2}
\bibinfo{author}{Bhargava, R.}, \bibinfo{author}{Khatri, C.},
  \bibinfo{year}{1981}.
\newblock \bibinfo{title}{The distribution of product of independent beta
  random variables with application to multivariate analysis}.
\newblock \bibinfo{journal}{Annals of the Institute of Statistical Mathematics}
  \bibinfo{volume}{33}, \bibinfo{pages}{287--296}.
\bibitem[{Bielak et~al.(2021)Bielak, Grzesiek, Janczura and
  Wy{\l}oma{\'n}ska}]{reso}
\bibinfo{author}{Bielak, {\L}.}, \bibinfo{author}{Grzesiek, A.},
  \bibinfo{author}{Janczura, J.}, \bibinfo{author}{Wy{\l}oma{\'n}ska, A.},
  \bibinfo{year}{2021}.
\newblock \bibinfo{title}{Market risk factors analysis for an international
  mining company. multi-dimensional heavy-tailed-based modelling}.
\newblock \bibinfo{journal}{Resources Policy} \bibinfo{volume}{74},
  \bibinfo{pages}{102308}.
\bibitem[{Brockwell and Davis(2016)}]{brockwell2016introduction}
\bibinfo{author}{Brockwell, P.J.}, \bibinfo{author}{Davis, R.A.},
  \bibinfo{year}{2016}.
\newblock \bibinfo{title}{Introduction to Time Series and Forecasting}.
\newblock \bibinfo{publisher}{Springer}.
\bibitem[{Cigizoglu and Bayazit(2000)}]{appl4}
\bibinfo{author}{Cigizoglu, H.K.}, \bibinfo{author}{Bayazit, M.},
  \bibinfo{year}{2000}.
\newblock \bibinfo{title}{A generalized seasonal model for flow duration
  curve}.
\newblock \bibinfo{journal}{Hydrological Processes} \bibinfo{volume}{14},
  \bibinfo{pages}{1053--1067}.
\bibitem[{Cochran(1934)}]{student_baza}
\bibinfo{author}{Cochran, W.G.}, \bibinfo{year}{1934}.
\newblock \bibinfo{title}{The distribution of quadratic forms in a normal
  system, with applications to the analysis of covariance}.
\newblock \bibinfo{journal}{Mathematical Proceedings of the Cambridge
  Philosophical Society} \bibinfo{volume}{30}, \bibinfo{pages}{178–191}.
\bibitem[{ENTSO-E()}]{ENTSOE}
ENTSO-E, \bibinfo{year}{2021}.
\newblock \bibinfo{title}{European association for the cooperation of
  transmission system operators ({TSO}s) for electricity}.
\newblock \bibinfo{note}{\url{}{https://transparency.entsoe.eu/}, accessed:
  2021-11-09}.
\bibitem[{Galambos and Simonelli(2004)}]{appl1}
\bibinfo{author}{Galambos, J.}, \bibinfo{author}{Simonelli, I.},
  \bibinfo{year}{2004}.
\newblock \bibinfo{title}{Products of Random Variables: Applications to
  Problems of Physics and to Arithmetical Functions (1st ed.)}.
\newblock \bibinfo{publisher}{Boca Raton: CRC Press}.
\bibitem[{Garg et~al.(2016)Garg, Sharma and Manohar}]{trapezoidal}
\bibinfo{author}{Garg, M.}, \bibinfo{author}{Sharma, A.},
  \bibinfo{author}{Manohar, P.}, \bibinfo{year}{2016}.
\newblock \bibinfo{title}{The distribution of the product of two independent
  generalized trapezoidal random variables}.
\newblock \bibinfo{journal}{Communications in Statistics - Theory and Methods}
  \bibinfo{volume}{45}, \bibinfo{pages}{6369 -- 6384}.
\bibitem[{Grzesiek et~al.(2020a)Grzesiek, Giri, Sundar and
  Wy{\l}oma{\'n}ska}]{ola22}
\bibinfo{author}{Grzesiek, A.}, \bibinfo{author}{Giri, P.},
  \bibinfo{author}{Sundar, S.}, \bibinfo{author}{Wy{\l}oma{\'n}ska, A.},
  \bibinfo{year}{2020}a.
\newblock \bibinfo{title}{Measures of cross-dependence for bidimensional
  periodic {AR}(1) model with alpha-stable distribution}.
\newblock \bibinfo{journal}{Journal of Time Series Analysis}
  \bibinfo{volume}{41}, \bibinfo{pages}{785--807}.
\bibitem[{Grzesiek et~al.(2020b)Grzesiek, Sikora, Teuerle and
  Wyłomańska}]{ola55}
\bibinfo{author}{Grzesiek, A.}, \bibinfo{author}{Sikora, G.},
  \bibinfo{author}{Teuerle, M.}, \bibinfo{author}{Wyłomańska, A.},
  \bibinfo{year}{2020}b.
\newblock \bibinfo{title}{Spatio-temporal dependence measures for bivariate
  {AR}(1) models with alpha-stable noise}.
\newblock \bibinfo{journal}{Journal of Time Series Analysis}
  \bibinfo{volume}{41}, \bibinfo{pages}{454--475}.
\bibitem[{Grzesiek et~al.(2019a)Grzesiek, Sundar and
  Wy{\l}oma{\'n}ska}]{grzesiekfloc}
\bibinfo{author}{Grzesiek, A.}, \bibinfo{author}{Sundar, S.},
  \bibinfo{author}{Wy{\l}oma{\'n}ska, A.}, \bibinfo{year}{2019}a.
\newblock \bibinfo{title}{Fractional lower order covariance-based estimator for
  bidimensional {AR(1)} model with stable distribution}.
\newblock \bibinfo{journal}{International Journal of Advances in Engineering
  Sciences and Applied Mathematics} \bibinfo{volume}{11},
  \bibinfo{pages}{217--229}.
\bibitem[{Grzesiek et~al.(2019b)Grzesiek, Teuerle and
  Wy{\l}oma{\'n}ska}]{ola11}
\bibinfo{author}{Grzesiek, A.}, \bibinfo{author}{Teuerle, M.},
  \bibinfo{author}{Wy{\l}oma{\'n}ska, A.}, \bibinfo{year}{2019}b.
\newblock \bibinfo{title}{Cross-codifference for bidimensional {VAR}(1) time
  series with infinite variance}.
\newblock \bibinfo{journal}{Communications in Statistics - Simulation and
  Computation} , \bibinfo{pages}{1--26,{ }}.
\bibitem[{Hansen(2003)}]{econ2}
\bibinfo{author}{Hansen, P.R.}, \bibinfo{year}{2003}.
\newblock \bibinfo{title}{Structural changes in the cointegrated vector
  autoregressive model}.
\newblock \bibinfo{journal}{Journal of Econometrics} \bibinfo{volume}{114},
  \bibinfo{pages}{261--295}.
\bibitem[{Homei(2019)}]{dirichlet}
\bibinfo{author}{Homei, H.}, \bibinfo{year}{2019}.
\newblock \bibinfo{title}{The stochastic linear combination of {Dirichlet}
  distributions}.
\newblock \bibinfo{journal}{Communications in Statistics - Theory and Methods}
  \bibinfo{volume}{50}, \bibinfo{pages}{2354 -- 2359}.
\bibitem[{Idrizi(2014)}]{pareto4}
\bibinfo{author}{Idrizi, L.}, \bibinfo{year}{2014}.
\newblock \bibinfo{title}{On the product and ratop of {Pareto} and
  {Kumaraswamy} random variables}.
\newblock \bibinfo{journal}{Mathematical Theory and Modeling}
  \bibinfo{volume}{4}, \bibinfo{pages}{137--146}.
\bibitem[{Johansen(2000)}]{econ1}
\bibinfo{author}{Johansen, S.}, \bibinfo{year}{2000}.
\newblock \bibinfo{title}{Modelling of cointegration in the vector
  autoregressive model}.
\newblock \bibinfo{journal}{Economic Modelling} \bibinfo{volume}{17},
  \bibinfo{pages}{359--373}.
\bibitem[{Lai and Balakrishnan(2009)}]{Lai_2009}
\bibinfo{author}{Lai, C.D.}, \bibinfo{author}{Balakrishnan, N.},
  \bibinfo{year}{2009}.
\newblock \bibinfo{title}{Continuous Bivariate Distributions}.
\newblock \bibinfo{publisher}{Springer New York}.
\bibitem[{Lee and Shih(2004)}]{lee2004product}
\bibinfo{author}{Lee, Y.J.}, \bibinfo{author}{Shih, H.H.},
  \bibinfo{year}{2004}.
\newblock \bibinfo{title}{The product formula of multiple {L{\'e}vy-It{\^o}}
  integrals}.
\newblock \bibinfo{journal}{Bulletin-Institute Of Mathematics Academia Sinica}
  \bibinfo{volume}{32}, \bibinfo{pages}{71--96}.
\bibitem[{Li et~al.(2020)Li, He and Blum}]{gauss2}
\bibinfo{author}{Li, Y.}, \bibinfo{author}{He, Q.}, \bibinfo{author}{Blum,
  R.S.}, \bibinfo{year}{2020}.
\newblock \bibinfo{title}{On the product of two correlated complex {Gaussian}
  random variables}.
\newblock \bibinfo{journal}{IEEE Signal Processing Letters}
  \bibinfo{volume}{27}, \bibinfo{pages}{16--20}.
\bibitem[{L{\"u}tkepohl(1985)}]{lut1}
\bibinfo{author}{L{\"u}tkepohl, H.}, \bibinfo{year}{1985}.
\newblock \bibinfo{title}{Comparison of criteria for estimating the order of a
  vector autoregressive process}.
\newblock \bibinfo{journal}{Journal of Time Series Analysis}
  \bibinfo{volume}{6}, \bibinfo{pages}{35--52}.
\bibitem[{Ly et~al.(2019)Ly, Pho, Ly and Wong}]{appl5}
\bibinfo{author}{Ly, S.}, \bibinfo{author}{Pho, K.H.}, \bibinfo{author}{Ly,
  S.}, \bibinfo{author}{Wong, W.K.}, \bibinfo{year}{2019}.
\newblock \bibinfo{title}{Determining distribution for the product of random
  variables by using copulas}.
\newblock \bibinfo{journal}{Risks} \bibinfo{volume}{7}.
\bibitem[{M.~Ahsanullah(2014)}]{Ahsanullah_2014}
\bibinfo{author}{M.~Ahsanullah, B. M. Golam~Kibria, M.S.},
  \bibinfo{year}{2014}.
\newblock \bibinfo{title}{Normal and Student’s t Distributions and Their
  Applications}.
\newblock \bibinfo{publisher}{Atlantis Press}.
\bibitem[{Malik and Trudel(1986)}]{exponential}
\bibinfo{author}{Malik, H.J.}, \bibinfo{author}{Trudel, R.},
  \bibinfo{year}{1986}.
\newblock \bibinfo{title}{Probability density function of the product and
  quotient of two correlated exponential random variables}.
\newblock \bibinfo{journal}{Canadian Mathematical Bulletin}
  \bibinfo{volume}{29}, \bibinfo{pages}{413–418}.
\bibitem[{Nadarajah(2005)}]{eliptically}
\bibinfo{author}{Nadarajah, S.}, \bibinfo{year}{2005}.
\newblock \bibinfo{title}{On the product xy for some elliptically symmetric
  distributions}.
\newblock \bibinfo{journal}{Statistics \& Probability Letters}
  \bibinfo{volume}{75}, \bibinfo{pages}{67--75}.
\bibitem[{Nadarajah(2008a)}]{logistic}
\bibinfo{author}{Nadarajah, S.}, \bibinfo{year}{2008}a.
\newblock \bibinfo{title}{Exact distribution of the product of two or more
  logistic random variables}.
\newblock \bibinfo{journal}{Methodology and Computing in Applied Probability}
  \bibinfo{volume}{11}, \bibinfo{pages}{651--660}.
\bibitem[{Nadarajah(2008b)}]{pareto2}
\bibinfo{author}{Nadarajah, S.}, \bibinfo{year}{2008}b.
\newblock \bibinfo{title}{On the product of generalized pareto random
  variables}.
\newblock \bibinfo{journal}{Applied Economics Letters} \bibinfo{volume}{15},
  \bibinfo{pages}{253 -- 259}.
\bibitem[{Nadarajah(2008c)}]{pearson}
\bibinfo{author}{Nadarajah, S.}, \bibinfo{year}{2008}c.
\newblock \bibinfo{title}{Some algebra for {Pearson} type vii random
  variables}.
\newblock \bibinfo{journal}{Bulletin of The Korean Mathematical Society}
  \bibinfo{volume}{45}, \bibinfo{pages}{339--353}.
\bibitem[{Nadarajah(2010)}]{pareto_gama}
\bibinfo{author}{Nadarajah, S.}, \bibinfo{year}{2010}.
\newblock \bibinfo{title}{Sum, product and ratio of pareto and gamma
  variables}.
\newblock \bibinfo{journal}{Journal of Statistical Computation and Simulation}
  \bibinfo{volume}{80}, \bibinfo{pages}{1071 -- 1082}.
\bibitem[{Nadarajah and Dey(2006)}]{student_t}
\bibinfo{author}{Nadarajah, S.}, \bibinfo{author}{Dey, D.K.},
  \bibinfo{year}{2006}.
\newblock \bibinfo{title}{On the product and ratio of t random variables}.
\newblock \bibinfo{journal}{Applied Mathematics Letters} \bibinfo{volume}{19},
  \bibinfo{pages}{45--55}.
\bibitem[{Nadarajah and Kotz(2005a)}]{normal_laplace}
\bibinfo{author}{Nadarajah, S.}, \bibinfo{author}{Kotz, S.},
  \bibinfo{year}{2005}a.
\newblock \bibinfo{title}{A note on the product of normal and laplace random
  variables}.
\newblock \bibinfo{journal}{Brazilian Journal of Probability and Statistics}
  \bibinfo{volume}{19}, \bibinfo{pages}{33--38}.
\bibitem[{Nadarajah and Kotz(2005b)}]{gama_beta}
\bibinfo{author}{Nadarajah, S.}, \bibinfo{author}{Kotz, S.},
  \bibinfo{year}{2005}b.
\newblock \bibinfo{title}{On the product and ratio of gamma and beta random
  variables}.
\newblock \bibinfo{journal}{Allgemeines Statistisches Archiv}
  \bibinfo{volume}{89}, \bibinfo{pages}{435--449}.
\bibitem[{Nadarajah and Kotz(2006)}]{gama_weibul}
\bibinfo{author}{Nadarajah, S.}, \bibinfo{author}{Kotz, S.},
  \bibinfo{year}{2006}.
\newblock \bibinfo{title}{On the product and ratio of gamma and {Weibull}
  random variables}.
\newblock \bibinfo{journal}{Econometric Theory} \bibinfo{volume}{22},
  \bibinfo{pages}{338--344}.
\bibitem[{Nadarajah and Kotz(2008)}]{appl10}
\bibinfo{author}{Nadarajah, S.}, \bibinfo{author}{Kotz, S.},
  \bibinfo{year}{2008}.
\newblock \bibinfo{title}{Sociological models based on {Fr{\'e}chet} random
  variables}.
\newblock \bibinfo{journal}{Quality \& Quantity} \bibinfo{volume}{42},
  \bibinfo{pages}{89--95}.
\bibitem[{Nadarajah and Kotz(2011)}]{gauss_laplace2}
\bibinfo{author}{Nadarajah, S.}, \bibinfo{author}{Kotz, S.},
  \bibinfo{year}{2011}.
\newblock \bibinfo{title}{On the linear combination, product and ratio of
  normal and {Laplace} random variables}.
\newblock \bibinfo{journal}{J. Frankl. Inst.} \bibinfo{volume}{348},
  \bibinfo{pages}{810--822}.
\bibitem[{Nadarajah and Kotz(2016)}]{pearson_2}
\bibinfo{author}{Nadarajah, S.}, \bibinfo{author}{Kotz, S.},
  \bibinfo{year}{2016}.
\newblock \bibinfo{title}{On the product and ratio of {Pearson} type vii and
  {Laplace} random variables}.
\newblock \bibinfo{journal}{Austrian Journal of Statistics}
  \bibinfo{volume}{34}, \bibinfo{pages}{11--23}.
\bibitem[{Podolski(1972)}]{appl2}
\bibinfo{author}{Podolski, H.}, \bibinfo{year}{1972}.
\newblock \bibinfo{title}{The distribution of a product of n independent random
  variables with generalized gamma distribution}.
\newblock \bibinfo{journal}{Demonstratio Mathematica} \bibinfo{volume}{4},
  \bibinfo{pages}{119 -- 124}.
\bibitem[{Roussas(2015)}]{Roussas_2015}
\bibinfo{author}{Roussas, G.G.}, \bibinfo{year}{2015}.
\newblock \bibinfo{title}{Joint and conditional p.d.f.'s, conditional
  expectation and variance, moment generating function, covariance, and
  correlation coefficient}.
\newblock \bibinfo{journal}{In: An Introduction to Probability and Statistical
  Inference} , \bibinfo{pages}{135--186}.
\bibitem[{Russo and Vallois(1998)}]{russo1998product}
\bibinfo{author}{Russo, F.}, \bibinfo{author}{Vallois, P.},
  \bibinfo{year}{1998}.
\newblock \bibinfo{title}{Product of two multiple stochastic integrals with
  respect to a normal martingale}.
\newblock \bibinfo{journal}{Stochastic processes and their applications}
  \bibinfo{volume}{73}, \bibinfo{pages}{47--68}.
\bibitem[{Saikkonen and L{\"u}tkepohl(2000)}]{lut2}
\bibinfo{author}{Saikkonen, P.}, \bibinfo{author}{L{\"u}tkepohl, H.},
  \bibinfo{year}{2000}.
\newblock \bibinfo{title}{Trend adjustment prior to testing for the
  cointegrating rank of a vector autoregressive process}.
\newblock \bibinfo{journal}{Journal of Time Series Analysis}
  \bibinfo{volume}{21}, \bibinfo{pages}{435--456}.
\bibitem[{Salo et~al.(2006)Salo, El-Sallabi and Vainikainen}]{appl6}
\bibinfo{author}{Salo, J.}, \bibinfo{author}{El-Sallabi, H.},
  \bibinfo{author}{Vainikainen, P.}, \bibinfo{year}{2006}.
\newblock \bibinfo{title}{The distribution of the product of independent
  {Rayleigh} random variables}.
\newblock \bibinfo{journal}{IEEE Transactions on Antennas and Propagation}
  \bibinfo{volume}{54}, \bibinfo{pages}{639--643}.
\bibitem[{Seijas-Mac{\'i}as and Oliveira(2012)}]{gauss1}
\bibinfo{author}{Seijas-Mac{\'i}as, A.}, \bibinfo{author}{Oliveira, A.},
  \bibinfo{year}{2012}.
\newblock \bibinfo{title}{An approach to distribution of the product of two
  normal variables}.
\newblock \bibinfo{journal}{Discussiones Mathematicae Probability and
  Statistics} \bibinfo{volume}{32}, \bibinfo{pages}{87--99}.
\bibitem[{Shakil and Kibria(2007)}]{maxwel}
\bibinfo{author}{Shakil, M.}, \bibinfo{author}{Kibria, B.M.},
  \bibinfo{year}{2007}.
\newblock \bibinfo{title}{On the product of {Maxwell} and {Rice} random
  variables}.
\newblock \bibinfo{journal}{Journal of Modern Applied Statistical Methods}
  \bibinfo{volume}{6}, \bibinfo{pages}{19}.
\bibitem[{Sheskin(2011)}]{tests_book}
\bibinfo{author}{Sheskin, D.J.}, \bibinfo{year}{2011}.
\newblock \bibinfo{title}{Handbook of Parametric and Nonparametric Statistical
  Procedures (5th ed.)}.
\newblock \bibinfo{publisher}{Chapman and Hall/CRC}, \bibinfo{address}{Boca
  Raton}.
\bibitem[{Tang and Gupta(1984)}]{beta}
\bibinfo{author}{Tang, J.}, \bibinfo{author}{Gupta, A.}, \bibinfo{year}{1984}.
\newblock \bibinfo{title}{On the distribution of the product of independent
  beta random variables}.
\newblock \bibinfo{journal}{Statistics \& Probability Letters}
  \bibinfo{volume}{2}, \bibinfo{pages}{165--168}.
\bibitem[{Tella and Geiss(2020)}]{tella}
\bibinfo{author}{Tella, P.D.}, \bibinfo{author}{Geiss, C.},
  \bibinfo{year}{2020}.
\newblock \bibinfo{title}{Product and moment formulas for iterated stochastic
  integrals (associated with {L}\'evy processes)}.
\newblock \bibinfo{journal}{Stochastics} \bibinfo{volume}{92},
  \bibinfo{pages}{969--1004}.
\bibitem[{Wecker(1978)}]{wecker}
\bibinfo{author}{Wecker, W.E.}, \bibinfo{year}{1978}.
\newblock \bibinfo{title}{A note on the time series which is the product of two
  stationary time series}.
\newblock \bibinfo{journal}{Stochastic Processes and their Applications}
  \bibinfo{volume}{8}, \bibinfo{pages}{153--157}.
\bibitem[{Weron(2006)}]{RWeron_energy}
\bibinfo{author}{Weron, R.}, \bibinfo{year}{2006}.
\newblock \bibinfo{title}{Modeling and forecasting electricity loads and
  prices: a statistical approach}.
\newblock Wiley Finance Series, \bibinfo{publisher}{John Wiley \& Sons},
  \bibinfo{address}{Chichester}.
\bibitem[{Williams(1992)}]{matrix}
\bibinfo{author}{Williams, K.S.}, \bibinfo{year}{1992}.
\newblock \bibinfo{title}{The nth power of a $2 \times 2$ matrix}.
\newblock \bibinfo{journal}{Mathematics Magazine} \bibinfo{volume}{65(5)},
  \bibinfo{pages}{336--336}.
\bibitem[{Wilson and Toumi(2005)}]{appl3}
\bibinfo{author}{Wilson, P.S.}, \bibinfo{author}{Toumi, R.},
  \bibinfo{year}{2005}.
\newblock \bibinfo{title}{A fundamental probability distribution for heavy
  rainfall}.
\newblock \bibinfo{journal}{Geophysical Research Letters} \bibinfo{volume}{32}.
\bibitem[{Yang and Wang(2013)}]{appl7}
\bibinfo{author}{Yang, Y.}, \bibinfo{author}{Wang, Y.}, \bibinfo{year}{2013}.
\newblock \bibinfo{title}{Tail behavior of the product of two dependent random
  variables with applications to risk theory}.
\newblock \bibinfo{journal}{Extremes} \bibinfo{volume}{16},
  \bibinfo{pages}{55--74}.
\bibitem[{Zivot and Wang(2006)}]{fin1}
\bibinfo{author}{Zivot, E.}, \bibinfo{author}{Wang, J.}, \bibinfo{year}{2006}.
\newblock \bibinfo{title}{Vector autoregressive models for multivariate time
  series}, in: \bibinfo{booktitle}{Modeling Financial Time Series with S-PLUS}.
  \bibinfo{publisher}{Springer}, \bibinfo{address}{New York}, pp.
  \bibinfo{pages}{385--429}.

\end{thebibliography}
\section*{Appendix}\label{sec:appendix}
\subsection*{\textbf{Bivariate Gaussian distribution}}
The bivariate Gaussian distributed random vector $(Z_1,Z_2)$ has the following PDF \citep{Roussas_2015} 
\begin{equation}\label{pdf_gaussian_baza}
       f_{Z_1,Z_2} (z_1,z_2) = \frac{ \exp\Bigg\{ -\frac{1}{2(1-\rho^2)} \left[  \frac{(z_1-\mu_{Z,1})^2}{\sigma_{Z,1}^2} 
        - 2\rho \left(\frac{z_1-\mu_{Z,1}}{\sigma_{Z,1}}\right)\left(\frac{z_2-\mu_{Z,2}}{\sigma_{Z,2}}\right)
        +\frac{(z_2-\mu_{Z,2})^2}{\sigma_{Z,2}^2}\right]\Bigg\}}{2 \pi \sigma_{Z,1} \sigma_{Z,2} \sqrt{1-\rho^2}},~~z_1,z_2\in \mathbb{R}
\end{equation}
where $\rho\in (-1,1)$ is the correlation coefficient between random variables $Z_1$ and $Z_2$ (denoted in the main text as $\rho_Z$); $\mu_{Z,1},\mu_{Z,2}\in \mathbb{R}$ are the corresponding expected values, while $\sigma_{Z,1}^2, \sigma_{Z,2}^2>0$ are the corresponding variances. When $\rho=0$, the PDF of the random vector $(Z_1,Z_2)$ is just the product of the PDFs of the Gaussian distributed random variables. 

\subsection*{\textbf{Bivariate Student's t distribution}}
The bivariate Student's t distributed random vector $(Z_1,Z_2)$ is constructed as follows. Let us assume that $(N_1,N_2)$ is the bivariate Gaussian vector defined by the PDF  in Eq. (\ref{pdf_gaussian_baza})  with expected values equal to zero, unit variances and $\rho\in (-1,1)$ being its correlation coefficient. Moreover, let $\chi^2$ be the one-dimensional random variable with chi-square distribution with $\eta>0$ degrees of freedom and assume that  $(N_1,N_2)$ and $\chi^2$ are independent. Then the random vector defined as
\begin{eqnarray}
(Z_1,Z_2)=\frac{(N_1,N_2)}{\sqrt{\chi^2/\eta}}
\end{eqnarray}
has a bivariate Student's t distribution with $\eta$ degrees of freedom and its PDF is given by \citep{Lai_2009}
\begin{equation}\label{student1}
    f_{Z_1,Z_2}(z_1,z_2) = \frac{1}{2\pi \sqrt{1-\rho^2}}\left[ 1 + \frac{z_1^2 - 2\rho z_1z_2 + z_2^2}{\eta(1-\rho^2)} \right]^{-\frac{\eta+2}{2}},~~z_1,z_2\in\mathbb{R}.
\end{equation}
The marginal random variables $Z_1$ and $Z_2$ have the one-dimensional Student's t distribution defined by the following PDF \citep{student_baza}
\begin{eqnarray}\label{student2}
f_{Z_1}(z_1)=\frac{\Gamma((\eta+1)/2)}{\sqrt{\eta\pi}\Gamma(\eta/2)}\left(1+\frac{z_1^2}{\eta}\right)^{-\frac{\eta+1}{2}},~z_1\in\mathbb{R},
\end{eqnarray}
where $\Gamma(\cdot)$ is the gamma function, i.e. $\Gamma(\alpha) = \int_{0}^{\infty} t^{\alpha-1} e^{-t} dt$ for $\alpha$ such that $\text{Re}(\alpha)>0$. Note that the number of degrees of freedom, $\eta$, is equal for both marginal variables. It is worth highlighting that the correlation $\rho_Z$ between the random variables $Z_1$ and $Z_2$ is equal to the parameter $\rho$. However, its zero value is not equivalent to the independence of the random variables $Z_1$ and $Z_2$, since in that case the PDF of a random vector $(Z_1,Z_2)$ (see Eq. (\ref{student1})) is not a product of the PDFs of the marginal distributions, see Eq. (\ref{student2}). Hence, if $Z_1$ and $Z_2$ are independent, the PDF of the random vector is given by
\begin{eqnarray}\label{student_ind}
 f_{Z_1,Z_2}(z_1,z_2)=\frac{\Gamma((\eta_{Z,1}+1)/2)\Gamma((\eta_{Z,2}+1)/2)}{\sqrt{\eta_{Z,1}\eta_{Z,2}}\pi\Gamma(\eta_{Z,1}/2)\Gamma(\eta_{Z,2}/2)} \left(1+\frac{z_{Z,1}^2}{\eta_{Z,1}}\right)^{-\frac{\eta_{Z,1}+1}{2}}\left(1+\frac{z_2^2}{\eta_{Z,2}}\right)^{-\frac{\eta_{Z,2}+1}{2}},~z_1, z_2\in\mathbb{R},
\end{eqnarray}
where $\eta_{Z,1}>0, \eta_{Z,2}>0$ are the degrees of freedom parameters of $Z_1$ and $Z_2$, respectively.

The Student's t distribution defined in (\ref{student2}) has zero mean and variance equal to $\sigma^2_{Z,1}=\frac{\eta}{\eta-2}$. It can be generalized to the Student's t location-scale distribution by applying the following transformation $Z_{(\mu,\lambda)} := \mu + \lambda Z$, where $Z$ is Student's t distributed. It yields a three parameter $(\mu,\lambda,\eta)$ distribution, with $\mu$ being the shift parameter, $\lambda>0$ the scale parameter and $\eta>0$ the degrees of freedom. The variance of the Student's t location-scale random variable is equal to $\sigma^2_{Z_{(\mu,\lambda)}}=\lambda^2 \frac{\eta}{\eta-2}$.
\end{document}